# Molecular genetics and mid-career economic mobility

Paul Minard
Heritage College
Gatineau, QC
pminard@cegep-heritage.qc.ca
August 2022

**Abstract**

Reductions in the cost of genetic sequencing have enabled the construction of large datasets including both genetic and phenotypic data. Based on these datasets, polygenic scores (PGSs) summarizing an individual's genetic propensity for educational attainment have been constructed. It is by now well established that this PGS predicts wages, income, and occupational prestige and occupational mobility across generations. It is unknown whether a PGS for educational attainment can predict upward income and occupational mobility even within the peak earning years of an individual. Using data from the Wisconsin Longitudinal Study (WLS), I show that: (*i*) a PGS for educational attainment predicts wage, income and occupational prestige mobility between 1974 (when respondents were about 36 years of age) and 1992 (when respondents were about 53 years of age), conditional on 1974 values of these variables and a range of covariates; (ii) the effect is not mediated by parental socioeconomic status, is driven primarily by respondents with only a high school education, and is replicated in a within sibling-pair design; (*iii*) conditional on 1974 outcomes, higher PGS individuals surveyed in 1975 aspired to higher incomes and more prestigious jobs 10 years hence, an effect driven primarily by respondents with more than a high school education; (*iv*) throughout their employment history, high PGS individuals were more likely to undertake on the job training, and more likely to change job duties during tenure with an employer; and (*v*) though no more likely to change employers or industries during their careers, high PGS individuals were more likely in 1974 to be working in industries which would experience high wage growth in subsequent decades. These results contribute to our understanding of longitudinal inequality across careers and shed light on the sources of heterogeneity in responses to economic shocks and policy.

I am grateful to Coady Wing for helpful comments. Current draft available here.

1. **Introduction**

Evidence from twin and adoption studies has long indicated the relevance of genetics for economic outcomes of interest (e.g., Cesarini and Visscher, 2017; Sacerdote, 2007; Silventionen et al., 2020; Branigan, McCallum and Freese, 2013).  Reviewing 21 existing twin studies of the genetic heritability of income, Hyytinen et al. (2019) note broadly consistent results across countries estimating that about 40% of the variation in income is attributable to genetic factors, with little role for the shared environment, and a substantial role (about 50%) for the non-shared environment, findings which they replicate in a study of Finnish twins.

Admittedly, the robustness of estimates of genetic contributions to economic outcomes via twin studies depends on the defensibility of the assumption that the environments experienced by monozygotic twins are as similar as those experienced by dizygotic twins.  In recent decades, however, reductions in the cost of genetic sequencing have led to the genotyping of large samples which, when paired with phenotypic data, enable association of specific locations on the genome with traits of interest.  For instance, analyzing a sample of 1.1 million individuals, Lee et al. (2018) identify 1,271 loci of variation in the genome (single nucleotide polymorphisms (SNPs)) which are independently associated with educational attainment.  The authors find that a polygenic score (PGS), a summary measure of the additive effect of SNPs identified as significantly associated with educational attainment in a genome-wide association study (GWAS), weighted by SNP-specific effect size, is able to predict 11-13% of the variance in educational attainment out of sample, making it a better predictor than household income in childhood.  It is important to emphasize that these estimates derive from direct reading of the genes of unrelated individuals, and do not rely on the assumptions of traditional twin and adoption studies.

Several large longitudinal social science datasets now include polygenic scores for educational attainment for a subset of genotyped study participants, and have been employed in recent economic research.  Using data from the Health and Retirement Study (HRS), Barth, Papageorge and Thom(2020) show that an educational attainment PGS predicts household income and wealth at retirement, even after conditioning on realized education.  Papageorge and Thom (2020) also find an association between the PGS and labor earnings.  Sias, Starks and Turtle (2020) find that individuals with higher PGSs for educational attainment are more likely to invest in equity markets.  Using data from the Dunedin study, Belsky et al. (2016) also demonstrate higher incomes and educational attainment among those with higher PGSs, as well as greater facility with financial planning.  Given the transmission of genes from parents to children, researchers have also begun to assess the implications of genetics for our understanding of social stratification and mobility.  Belsky et al. (2016) show that mobility relative to socioeconomic origins is higher among those with a higher educational attainment PGS.  Across five longitudinal studies, Belsky et al. (2018) find that children with higher education attainment PGSs were more upwardly mobile than their parents in terms of realized educational attainment, occupational prestige, and wealth.

Databases pairing genetic data with longitudinal measures of economic outcomes enable assessment of whether the importance of genetic variation in accounting for phenotypic variation changes across the life course.  Two such databases are the aforementioned HRS and the Wisconsin Longitudinal Study (WLS), discussed in more detail below.  Respondents in these surveys were in their prime working years in the latter quarter of the 20[th] century, a period characterized by increasing college wage premiums and technological disruption in advanced economies.  If responses to technological shocks are characterized by substantial heterogeneity (Keane, Moffit and Runkle, 1988), human capital accumulation is of rising economic importance, and genetic endowments strongly predict both measured and unmeasured human capital accumulation, then genetic variation is an increasingly important source of heterogeneity in economic outcomes.  Papageorge and Thom (2020) find an increasing association between an educational attainment PGS, controlling for measured education, and labor market earnings after 1980, coincident with a period of technological change.  One interpretation of this result is that genes associated with measured educational attainment are, more conceptually, associated with a tendency to acquire human capital, and may help individuals master new technologies and acquire new skills, better withstanding the winds of skill-biased technological change.  If genes associated with measured education predict ongoing economic advancement throughout the life course, conditional on initial education and economic outcomes, this has important implications for our understanding of the sources of economic inequality, heterogeneity in responses to economic shocks, and heterogeneity across individuals in the impact of policies designed to help workers cope with skill biased technological change, such as retraining programs.  Although phenotypic measures of cognitive ability have already been shown to predict upward mobility in job complexity across a five-year period (Wilk et al. 1995; Wilk and Sackett, 1996; Judge et al. 2010), it is not known whether upward mobility in income and occupational prestige during an individual's working years can be predicted from genetics alone.  If mobility during one's career can be predicted by genetics, this also means that the heritability of economic outcomes varies across the life course.

I address these questions using data from the Wisconsin Longitudinal Study (WLS). Beginning in 1957, the WLS surveyed a representative sample of Wisconsin high school graduates. This initial survey and subsequent waves have obtained information on job history, income, educational attainment, and a wide variety of demographic and health metrics. Unlike other longitudinal datasets that incorporate PGSs, the WLS data measure wages, income, and occupational status at varied points in the lifecycle and not only at retirement. At the time of subsequent survey waves, these graduates were approximately 36 years old (in 1975) and approximately 53 years old (in 1992). The WLS data therefore contain varied measures of economic status on either end of this prime earnings period in the life cycle. I show in a sample of 5,414 genotyped graduates that conditional on age, gender and wages in 1974, individuals with a higher educational attainment PGS have higher wages in 1992. That is, for two same-sex individuals of the same age and with the same wage in 1974, the individual with the higher educational attainment PGS tended to have higher wages in 1992. The effect sizes are of practical significance, as a 1-standard deviation increase in the PGS predicts an additional $2,000 in annual wages in 1992 dollars, relative to 1974 wages. Similar results are found for income mobility, with a 1-standard deviation increase in the PGS predicting an additional $3,000 in annual income in 1992, relative to 1974 income. I also show that a higher education PGS predicts greater increases in occupational prestige according to two measures. These effects are robust to the inclusion of a control for realized education, in the form of a dummy variable indicating whether the respondent obtained more than a high school education. Interestingly, the magnitude of the effect of the PGS on wage and income mobility is higher among respondents with only a high school education (59% of respondents in 1975), whereas the changes in occupational prestige are driven by respondents with more than a high school education.

The WLS data have several features which enable extensions beyond these basic findings. In exploring the mechanisms behind this result, I find that high PGS respondents were more likely to take on the job training throughout their employment history, and more likely to change job duties during a spell with an employer. Although high PGS respondents were not more likely to change employers or industries during their careers, I present evidence that high PGS respondents selected themselves, in the early 1970s, into industries which would experience high median wage growth in the ensuing decades.

In 1975 respondents were surveyed on their aspirations for their occupations 10 years hence. I find that individuals with a higher educational attainment PGS were more aspirational about the prestige of their future occupation, conditional on the prestige of their current occupation. Though this result is robust to the inclusion of the realized education dummy noted above, the PGS-aspiration association is almost entirely driven by respondents with more than a high school education.

If parents transmit cultural and financial capital as well as their genes to their offspring, genetic associations could be an artefact of social class, particularly if assortative mating is widespread. Hence genetic effects are more credibly identified using a within-sibling design, which aims to robustly control for common family effects, including the family's socioeconomic status. Belsky et al. (2018) find that siblings with higher PGSs for educational attainment are indeed more upwardly mobile than their siblings with lower educational attainment. Importantly, the WLS includes data on a randomly selected sibling of graduate respondents. Graduates were surveyed in 1975 and 1992, and siblings were surveyed in 1977 and 1994. In a sibling-difference analysis using 1,978 sibling pairs, siblings with the higher rank in education PGS tended to be more upwardly mobile between the two survey waves in terms of wage income, conditioning on age, gender, and initial wage income. The magnitudes of these estimates are similar to those obtained in analyses of the graduate-only sample.

These results bear on literatures from both behavioral genetics and economics. A consistent finding in the behavioral genetics literature is that the heritability of many traits, including cognitive ability, increases with age (e.g., Bouchard, 2013; Haworth et al., 2010). As the twin literature has found limited effects of the common environment in variance decomposition analyses, this increase in heritability over time comes at the expense of the importance of the nonshared environment. This is consistent with a model of an individual's economic life course in which one's economic position at any point in time is a function of genetics and shocks from the nonshared environment. The increasing importance of genetics as one ages implies that the impact of most of these environmental shocks is temporary and non-cumulative. In time series econometrics terms, economic position reverts over time to the long-run genetic mean, and the nonshared environment contributes single-period shocks that, while they impact current economic position, do not have permanent effects and are uncorrelated. At earlier stages in life, the impact of negative (positive) environmental shocks can lead to an economic position below (above) one's long-run genetic mean, but as the balance between positive and negative shocks evens out across the life course, one rises (falls) to the lifetime mean predicted by the unchanging effects of the genes. I call this the "wash-out effect". This is consistent with theory and empirical findings in behavioral genetics that greater equalization in environments increases the heritability of outcomes. Though this has been demonstrated convincingly for society-wide equalizations in economic opportunity (e.g.,

Rimfeld et al., 2018), we can think of the increasingly mean-zero impacts of positive and negative shocks across the life course as a form of environmental equalization over time.

An alternative account of increasing heritability is that nonshared environmental effects are of lasting importance, but with age individuals are better positioned to actively choose their own environments and do so in a way that is non-random with respect to their genotype (Plomin et al. 1977; Scarr, 1996). Plomin et al. (1977) proposed three types of genotype-environment correlation: passive correlation, whereby, for instance, parents of high verbal ability pass on genes correlated with this trait and also provide an environment conducive to the development of this trait, perhaps driven by their own genotypes (a nature of nurture effect); reactive correlation, wherein a child's genotype elicits environmental responses from parents and other agents of socialization (a nurture of nature effect); and active correlation, wherein the child seeks out environments conducive to the development of phenotypes to which they are already genetically predisposed. Consider a person of high genetic cognitive ability whose childhood environment was below average in terms of developing this phenotypic trait. With age, individuals are increasingly better positioned to select into environments that are better suited to their genotype, whether these be peer groups, educational settings, or, as adults, occupations whose nature are more commensurate with their underlying genetic propensities. This predicts increasing convergence of phenotypes and genotypes over time, as people gradually select an environment which facilitates phenotypic development, conditional on genotype. In effect, in this model environmental shocks become progressively less random over time.

Yet another contributing factor to increasing heritability may be that beyond genetic nurture effects, parents prefer to equalize the opportunities of their children during the children's formative years. That is, they may devote more (fewer) resources to support the educational and economic attainment of their relatively less (more) able offspring. Such compensating effects are not captured by family fixed effects models and are part of the nonshared environment of children (Fletcher et al. 2020). As children age and enter the workforce, ongoing parental efforts to equalize economic advancement are likely less impactful, and the fade-out of this compensating activity amounts to an equalization in the nonshared environment unless the effect of these compensating investments has been permanent. Fletcher et al. (2020) find that among sibling pairs, an educational attainment PGS predicts educational attainment less well for the more highly ranked, in terms of PGS, of two siblings, in support of the compensation effect among adolescents and young adults. In contrast, I show below that there is no statistically significant interaction between a sibling's rank within their sibling-pair and their own PGS in predicting income mobility.

The economics literature has tended to address these questions in the context of models of dynamic complementarity between genetic endowments and environmental investments in skill formation. Muslimova et al. (2020) show that birth order, which is correlated with parental time investments in young children, interacts with educational attainment PGSs in predicting years of education acquired, a nurture of nature effect.  Houmark, Ronda and Rosholm (2020) employ a dataset including parental as well as offspring genes, separating direct offspring genetic effects, indirect effects of parental investments (a nurture of nature effect) and the effect of parental genes on these investments (a nature of nurture effect).  Their dependent variables are measures of children's development and behavior in early life.  They identify both increasing direct effects of children's genes over time and a reinforcing nurture of nature effect.

The dynamic complementarity research has tended to focus on early life investments, as these are presumed to be most impactful for later life outcomes (Cunha and Heckman, 2007; Almond, Currie and Duque, 2018), and Muslimova et al. (2020) note it is an important question whether investments made beyond the formative years can compensate for the absence of early life investments by parents.  Cunha and Heckman (2007) review literature finding that positive effects of early life behavioural and cognitive interventions among disadvantaged children fade out unless followed up by future investments, and that by the time individuals are applying for college, family income plays a minor role in determining college attendance relative to ability.  Hence, in their view, the importance of sustained early, ability-enhancing interventions.

Our finding that an educational mobility PGS predicts mid-career economic mobility, conditional on early career outcomes, poses some challenges for the recent economics literature, but is consistent with a wash-out effect, which claims that all interventions fade out over time, including the nature of nurture and nurture of nature effects identified, for example, by Houmark, Ronda and Rosholm (2020). Proponents of a wash out account would note that the economics literature has tended to assess the nature of nurture and nurture of nature effects in early childhood.  If genetics can predict economic mobility even across prime earning years, this argues against a permanent impact of early life environmental interventions.  Although our results do not preclude lasting effects of parental nature of nurture, it would seem likely these parental investments would have matured by the time respondents were 36 years of age. Our results show, however, that genetics predicts upward mobility even past this age. The "wash-out effect" and a self-directed nurture of nature (what Plomin et al. (1977) called "active correlation") are more consistent with these results. Indeed, I test for interaction effects between parents' years of education and occupational prestige at the time of the child's high school graduation (1957), and the respondent's educational attainment PGS in predicting mid career income mobility, and find no significant interaction. This is evidence that our results are not driven by early life nurturing of genetic nature, and are better interpreted as direct genetic effects, perhaps complemented by self-directed nurture of nature effects.

In extensions to our main results, I find evidence that while high PGS graduates were not more likely to change employers or industries between 1974 and 1992, they were more likely, in 1974, to work in industries which would subsequently see high wage growth. I also find that high PGS graduates were more likely to undertake job training in hopes of finding a better job in the future and were more likely to see their main job duties change while working for an employer. In support of the findings of Papageorge and Thom (2020), these results suggest possible means by which a genetically-driven tendency to acquire human capital has helped some workers better navigate the challenges of an era characterized by skill biased technological change. As such, they are also informative about the sources of heterogeneity in response to ameliorative retraining and employment programs.

This paper contributes to a growing body of literature documenting the relevance of molecular genetic variation in accounting for individuals' economic status and economic inequality. Though the social implications of this literature may seem profound, it is important to recognize that we are in the earliest stages of the integration of molecular genetic data into the social sciences, and indeed, into economic and social life. The polygenic scores (construction of which is discussed in detail below) are employed here are in fact noisy proxies for underlying, latent genetic propensities (Muslimova et al., 2020). As the sample sizes of the genome-wide association studies from which these scores are constructed increase, the accuracy of these proximate measures will also increase, and with it their predictive power of outcomes of interest to economic agents.

The remainder of this paper proceeds as follows. Section 2 describes both the economic and genetic data used in the paper, with a particular emphasis on the creation of polygenic scores. It also describes the paper's approach to identification for both the graduate and sibling-pair samples. Section 3 presents main results on economic mobility, followed by extensions to consider the mechanisms underlying these effects. Section 4 concludes with a discussion of the implications of these results.

## 2. Data and identification

### 2.1. Data description and economic variables

The Wisconsin Longitudinal Study began as a state-sponsored census of 1957 Wisconsin high school graduates.  In 1962, one third of these respondents (n=10,317) were randomly sampled and became the focal cohort of "graduates" for subsequent survey rounds.  Graduates were surveyed in 1964 (at approximately 25 years old), 1975 (36 years old), 1992-3 (53 years old) , between 2003 and 2007, and between 2010 and 2011 (72 years old).  In 1977, a randomly selected sibling of the graduate sample was also surveyed.  These siblings were surveyed again in 1993-4 and 2004-7.  The WLS is directly representative of Wisconsin's high school graduating class of 1957, and more broadly of white American high school graduates, roughly 66% of the US population in 1990-91 (Pearce and Parks, 2011).[1]  Though the initial survey wave focused on graduates' aspirations for their education, and later waves have added substantial health-related questions, each survey wave has collected data on wages, income, and occupational history, as well as aspirations for future employment.  In this study, I focus on annual wages, income from all sources, job aspirations and occupational prestige.  Mean values for these variables, separated by provision of a DNA sample to the WLS, are presented in Table 1.

### 2.2. Genetic data

Human beings differ at less than 1% of the approximately 3 billion base pairs in the genome.  At each base pair, people have one of two nucleotide pairs, either adenine-thymine (AT) or guanine-cytosine (GC).  As DNA is inherited from each parent, at each loci people have two sites of potential variation, and three possible outcomes: AT-AT, GC-GC, or AT-GC.  At a randomly selected location along the genome, most people will have the same nucleotide pairs.  An important form of genetic variation among people is a single nucleotide polymorphism (SNP), at which most people have the "major" base pair, e.g., AT-AT, but some individuals have a "minor" base pair, e.g., AT-GC.  As people have two copies at each locus, we can represent the number of minor base pairs at each SNP as a count variable with the values 0, 1, or 2.

---

[1] I employ the public, prepackaged complete datasets compiled by WLS staff.  Data are freely available with registration.  See https://www.ssc.wisc.edu/wlsresearch/data/.  See also Herd et al. (2014).  Full codebooks are available here: https://www.ssc.wisc.edu/wlsresearch/documentation/waves/.

Dramatic reductions in the cost of genomic sequencing have enabled the construction of large datasets containing both genetic and phenotypic/behavioural data. A typical Genome Wide Association Study (GWAS) aims to identify SNPs at which variation is correlated with phenotypic variation. Phenotypes $Y$ are regressed on $J$ observable SNPs in $J$ separate regressions of the following form:

$$Y_i = \mu X_i' + \beta_j SNP_{ij} + \epsilon_{ij},$$

where $SNP_{ij} \in \{0,1,2\}$ is a count variable indicating the number of minor base pairs at location SNP $J$. Separate regressions are run for each SNP as the number of SNPs observed will typically be higher than the number of observations.

The regressions produce a set of $J$ GWAS coefficients $\{\hat{\beta}\}_{j=1}^{J}$ and standard errors. As there are millions of SNPs in the human genome, adopting a traditional p=0.05 significance threshold will lead to many false positives. A further challenge to the assumption of independence is linkage disequilibrium, which results from genes on chromosomes being inherited in a "chunky" fashion. This means that nearby SNPs are not inherited independently of one another. A full review of the methods for adjusting significance thresholds is beyond the scope of this paper. For an introductory treatment, see Mills et al. (2020). For the adjustments performed in the GWAS used to construct the WLS polygenic scores for educational attainment, see Lee et al. (2018).

Recent GWASs have indicated that cognitive and behavioural traits are nearly universally polygenic, meaning that many genes account for the sum genetic component of phenotypic variation, each of these generally having a very small effect. Robust identification of SNPs associated with a trait therefore requires large sample sizes. A large, recent GWAS for educational attainment is Lee et al. (2018)[2]. Their study of 1.1 million individuals found 1,271 independent genome-wide significant SNPs. This large number of SNPs indicates that educational attainment is highly polygenic. Predicting phenotypic values of a polygenic trait based on any one SNP is therefore impossible, and this polygenicity explains the failure of early "candidate-gene" studies in economics to replicate.[3] Polygenicity also means that whether we are assessing the share of phenotypic variance explained by genetics, or predicting phenotypic values based on genotypes, a summary measure of genotypes at $J$ SNPs is desirable. This is the motivation for the creation of polygenic scores, which is a weighted sum of one's genotype at $J$ SNPs. An individual's polygenic score for a phenotypic trait is constructed by a linear model:

$$PGS_i = \sum_{j=1}^{J} W_j \, SNP_{i,j}$$

where $PGS_i$ is the polygenic score for, in our case, educational attainment for individual $i$, $W_j$ is the weight for $SNP_j$ as derived from the GWAS, and $SNP_{i,j}$ is individual $i$'s genotype at $SNP_j$. It is important to note that the GWAS weights, $W_j$, were computed on studies of individuals not included in the WLS data. This means that our analysis is entirely out of sample. It is also important to note that polygenic scores are linear combinations of SNPs and their GWAS-derived effect sizes. As such they do not account for dominance or interaction effects[45]. Furthermore, although SNPs are measured accurately, the GWASs used to construct the weights have finite sample sizes, meaning that a PGS is a noisy estimate of the underlying genetic propensity for educational attainment. This leads to attenuation in coefficients on our PGS variable (Muslimova et al., 2020).

---

[2] Although Okbay et al. (2022) have constructed an updated polygenic score index using DNA samples from over 3m individuals, the WLS data has not been updated to include this improved score.
[3] See reviews of this literature in Fletcher (2018) and Mills and Tropf (2020).

A further interpretational concern is that polygenic scores for educational attainment may reflect population stratification. Consider the "chopsticks gene" scenario as described by Hamer and Copeland (2011). Imagine a GWAS is conducted on a global population with skill in using chopsticks as a phenotype. SNPs are duly identified as correlated with skill in chopstick use. The effect of these SNPs should not be interpreted as causal, however, as chopstick use is culturally transmitted from parents to children of, in the main, east Asian ancestry. The SNPs naively identified in this GWAS will merely be those which most clearly delineate this ancestry group from others. It is standard practice, therefore, when using a polygenic score as a regressor, to also include the SNP principal components as regressors to correct for this population stratification (Price et al. 2006). I therefore include the first 10 principal components in all regressions using the educational attainment polygenic score. Substantively, this addresses the concern that the PGS is merely capturing differences in cultural practices across ethnicities that are relevant for educational attainment (Papageorge and Thom, 2020). Note that in within-family analyses, these principal components are redundant, but I retain them in the sibling-pair analysis below to facilitate comparison with results from the graduate sample. Relatedly, GWAS studies to date have been based on samples of mostly European ancestry, and polygenic scores thereby derived have been shown to have lower predictive power among non-European ancestry groups, owing to differing patterns of linkage disequilibrium. As 99.6% of the WLS sample is of European descent, this is not a major issue in the present study.

---

[4] Okbay et al (2022) show that across common SNPs associated with educational attainment, dominance effects are almost entirely absent. An intuitive demonstration of additivity is the finding that effect sizes at SNPs are linear across increases in the number of minor alleles. That is, the effect size of moving from 0 to 1 minor alleles is similar to that of moving from 1-2 minor alleles. If dominance effects were widespread, effects would be concentrated in the move from 0-1 minor alleles at a locus.

[5] For a recent discussion of approaches to polygenic score construction, see Wang et al. (2022).

A subset of WLS respondents provided a DNA sample in the 2010-2011 collection wave. Samples were obtained from 5,414 graduates and 3,095 siblings. Polygenic scores (PGSs) for educational attainment were constructed for these individuals on the basis of the aforementioned Lee et al. (2018) GWAS. The WLS provides a dataset including these polygenic scores, which I link to the main dataset via the respondent's public ID. Though the polygenic score variable (hereafter PGS) is normally distributed, as predicted by Fisher's (1919) infinitesimal model, I standardize this variable so that it is mean centred at zero with a standard deviation of one. For the graduate-only models, this standardization is done among graduates only. This leads to a sample of 5,414 graduates for whom we have genetic data. For the within-sibling models, standardization is performed in the full sample of graduates and siblings. Our resulting sample has 1,978 sibling pairs.

About 52% of graduate respondents provided sample provided a DNA sample. Table 1 is a balance table comparing means of key variables among the genotyped sample (column 1) and the non-genotyped sample (column 2). Column (3) indicates the difference in means and assesses the statistical significance of the difference. The genotyped sample earned higher incomes and had higher occupational prestige scores in both survey waves. If higher PGSs are associated with economic attainment, and economic attainment is associated with participation, then individuals with low PGSs who are genotyped will tend to be those with higher economic attainments. This form of selection bias would attenuate positive associations between the educational attainment PGS and economic outcomes.

Table 1. Balance table, key variables by DNA sample provision

| Variable | (1) DNA | (2) No DNA | (3) Difference |
|---|---|---|---|
| Male | 0.477 | 0.491 | 0.014 |
| Age in 1974 | 38.862 | 38.820 | -0.042*** |
| H.S. degree only | 0.595 | 0.584 | -0.011 |
| Wage earnings, 1974 | 23,882.016 | 22,549.090 | -1,332.926** |
| Total earnings, 1974 | 26,139.164 | 24,186.592 | -1,952.572*** |
| Duncan prestige score, 1974 | 487.274 | 461.727 | -25.547*** |
| Siegel prestige score, 1974 | 442.917 | 425.939 | -16.977*** |
| Wage earnings, 1992 | 33,265.809 | 28,948.096 | -4,317.712*** |
| Total earnings, 1992 | 40,247.953 | 32,420.291 | -7,827.661*** |
| Duncan prestige score, 1992 | 493.342 | 460.681 | -32.661*** |
| Siegel prestige score, 1992 | 444.311 | 422.162 | -22.150*** |
| Observations | 5,414 | 4,903 | 10,317 |

### 2.3. Identification

#### 2.3.1. Graduate sample

I first assess the impact of the educational attainment PGS on economic mobility within the life course among the sample of graduates, using a lagged dependent variable model:

$$Y_{i,1992} = \theta Y_{i,1975} + \mu X'_i + \beta PGS_i + \varepsilon_i$$

where, $Y_{i,1992}$ is, in successive models, the graduate's 1992 wage, income or occupational prestige; $Y_{i,1975}$ is the value of this same variable in 1975; $X'_i$ are controls for gender, age, and the first 10 principal components. In some specifications, a dummy variable for having completed more than a high school education are also included in $X'_i$; $PGS_i$ is the graduate's standardized polygenic score for educational attainment.

In a second analysis, I assess whether the PGS can predict graduates' aspirations for their job in 10 years' time, conditional on current job as measured in the 1975 survey. I estimate the impact of the PGS on the prestige score of the aspired job via the following model:

$$Y_{i,1985} = \theta Y_{i,1975} + \mu X'_i + \beta PGS_i + \varepsilon_i$$

where $Y_{i,1985}$ is one of two operationalizations of prestige for the job aspired to 10 years in the future, as measured during the 1975 survey wave; $Y_{i1975}$ is the same operationalization of prestige for the graduate's then current job; $X'_i$ are the same set of controls noted above; and $PGS_i$ is the graduate's standardized polygenic score for educational attainment.

### 2.3.2.   Sibling pair sample

As noted above, concerns have been raised about whether the parameter $\beta$ in the above regressions has a causal interpretation, since if social advantage is transmitted culturally by parents to children along with DNA, then SNPs tagged as significant in GWAS studies may simply reflect social class stratification.  Although the inclusion of principal components mitigates against this concern, and the known role of the genes tagged by the Lee et al. (2018) GWAS in neurophysiological functioning suggests direct genetic effects, within-family designs are widely viewed as the preferred strategy for identifying the causal effects of SNP-level variation.  Indeed, as noted by Houmark et al. (2020), offspring genotypes are not exogenous to parental genes, and these parental genes impact the environment in which the offspring is raised (a nature of nurture effect).  Estimates derived from sibling-pair designs are robust to bias from the shared environment, including shared environmental effects driven by parental genotypes.  The WLS data enables a within-family design as a random sibling of graduates were identified and surveyed in the 1977 survey wave, creating a total of 10,270 sibling pairs.  Genetic data were obtained for both siblings in 1,978 of these pairs, and this is our sample for the within-family analysis.  I introduce family fixed effects to estimate the causal impact of higher educational attainment polygenic scores on within life-course mobility via the following model:

$$Y_{is,1992} = \theta Y_{is,1974} + \mu X'_i + \alpha F_s + \beta PGS_{is} + \varepsilon_{is}$$

where $Y_{is,1992}$ is the wage, in levels and logs in subsequent specifications, of individual $i$ in sibling pair $s$ in 1992; $Y_{is,1974}$ is the equivalent value in 1974; $X'_i$ are the individual-level controls described above; $F_s$ is a sibling-pair fixed effect; $PGS_{is}$ is, in subsequent specifications, is the respondent's educational PGS and a dummy variable equal to 1 for the higher ranked sibling.  Graduate respondents without siblings are omitted from this analysis.

Considerable variation in PGSs remains within sibling pairs.  Siblings on average share 50% of their genes, and in expectation their true PGS is an average of parental PGSs.  However, though thousands of SNPs are associated with education, the strength of these associations, $W_j$, is not constant across all SNPs. Therefore, some siblings will be fortunate enough to inherit the most significant SNPs with positive effects. Figure 1 plots the PGS of siblings (Y-axis) against the PGS of the corresponding graduate (x-axis). Though a strong association is apparent, considerable differences among siblings remain.

**Figure 1**. Association between graduate and sibling educational attainment PGSs

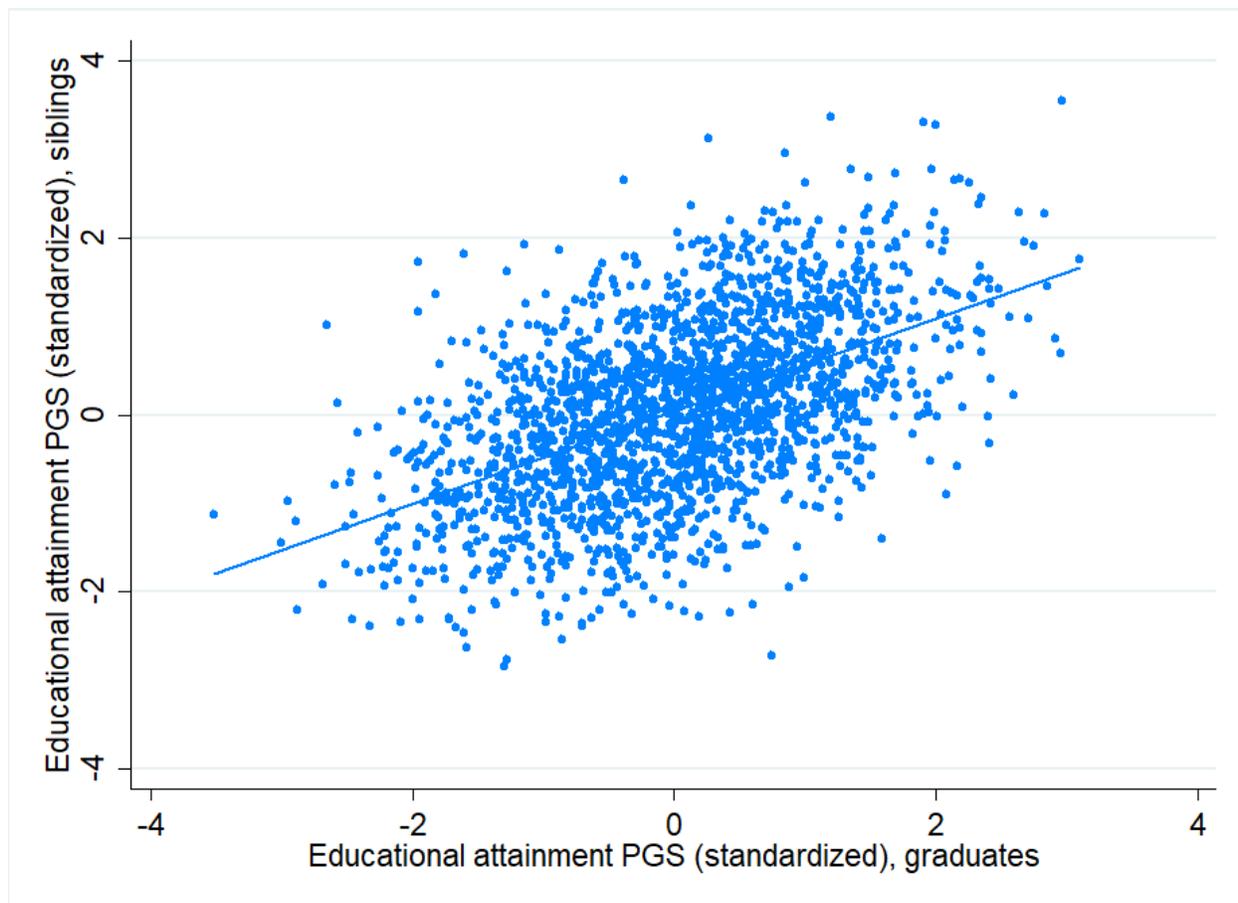

## 3. Results

### 3.1. Educational attainment PGS and realized education

Before turning to the main results, I first assess our educational attainment PGS as a predictor of realized education in the WLS data. Figure 2 shows a scatterplot of values of the WLS's "equivalent years of education" variable by values of the PGS. A clearly positive association is shown, higher values of the PGS predicting more years of education. The figure is overlayed with a kernel density plot of the standardized PGS variable, which is approximately normally distributed.

**Figure 2**. Scatterplot of years of education and educational attainment PGS; kernel density plot of PGS

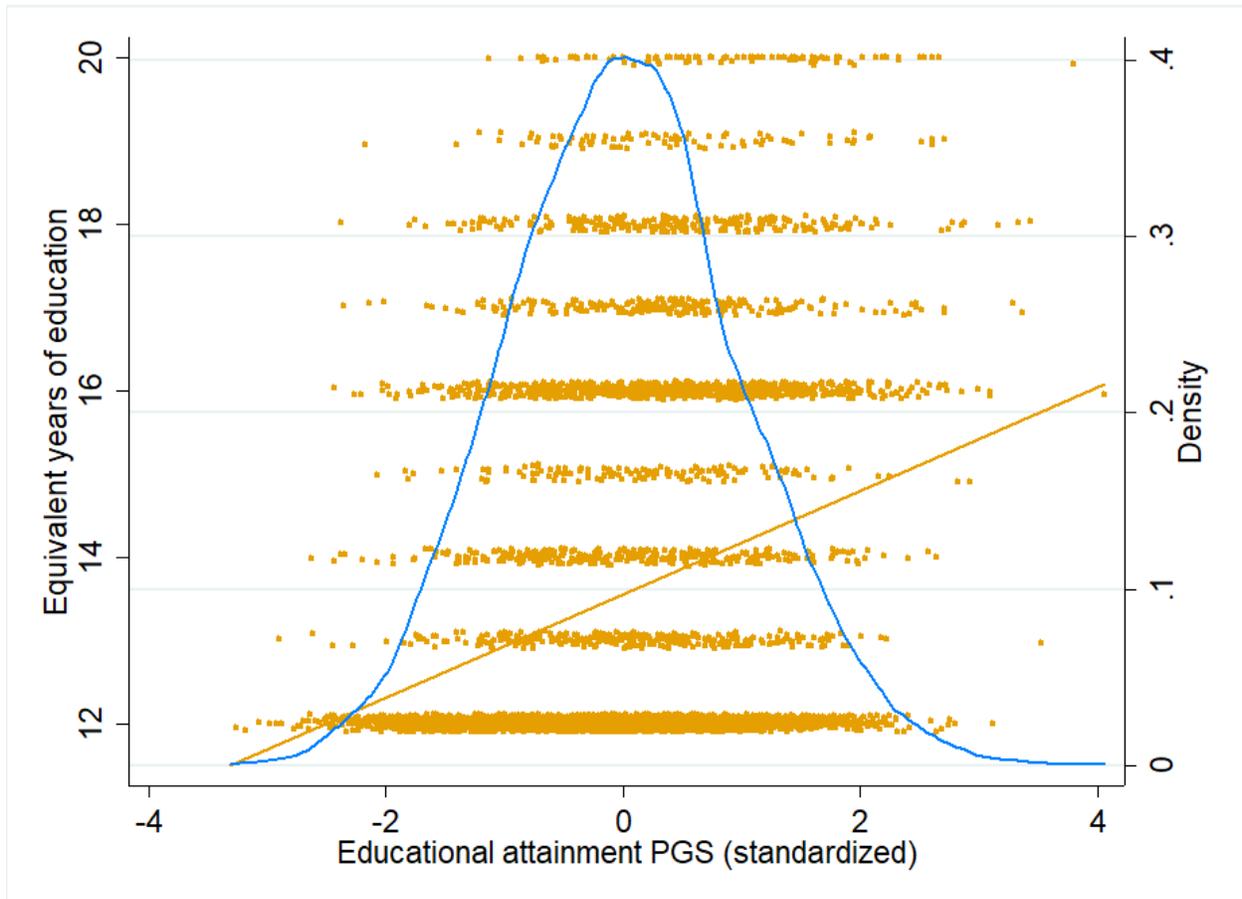

Figure 3 compares the distribution of the PGS variable across respondents with a high school education (solid black line) and with more than a high school education (dashed grey line). The distribution is clearly shifted to the right among those with more than a high school education. The difference in means is approximately 0.5 standard deviations.

**Figure 3.** Density plots of educational attainment PGS, by education

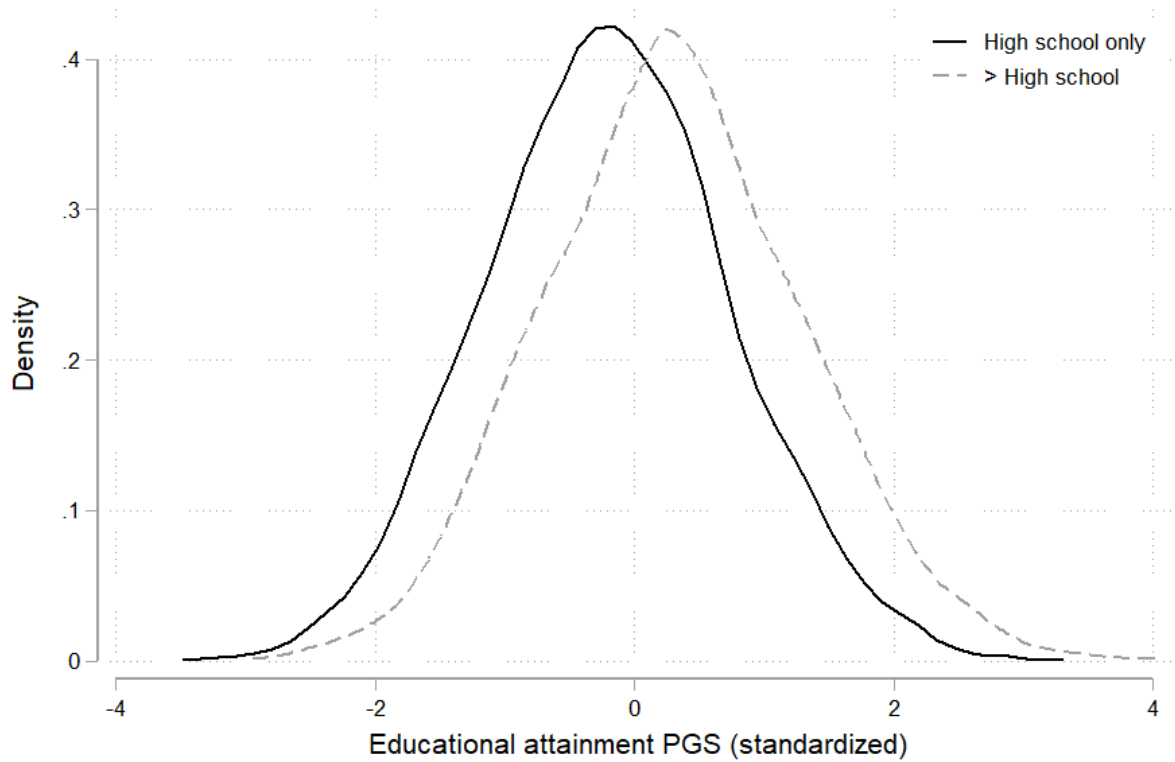

### 3.2. Main results
#### 3.2.1. Graduate sample

I first show descriptively that rankings in the educational attainment PGS predict income mobility without conditioning on covariates. Figure 4 compares median income in 1974 and 1992 across four quartiles of the PGS variable. Not only does the PGS predict median income in both time periods, but it is clear that the top quartile saw particularly large gains between 1974 and 1992.

**Figure 4**. Unconditional income mobility, 1974-1992, by PGS quartile

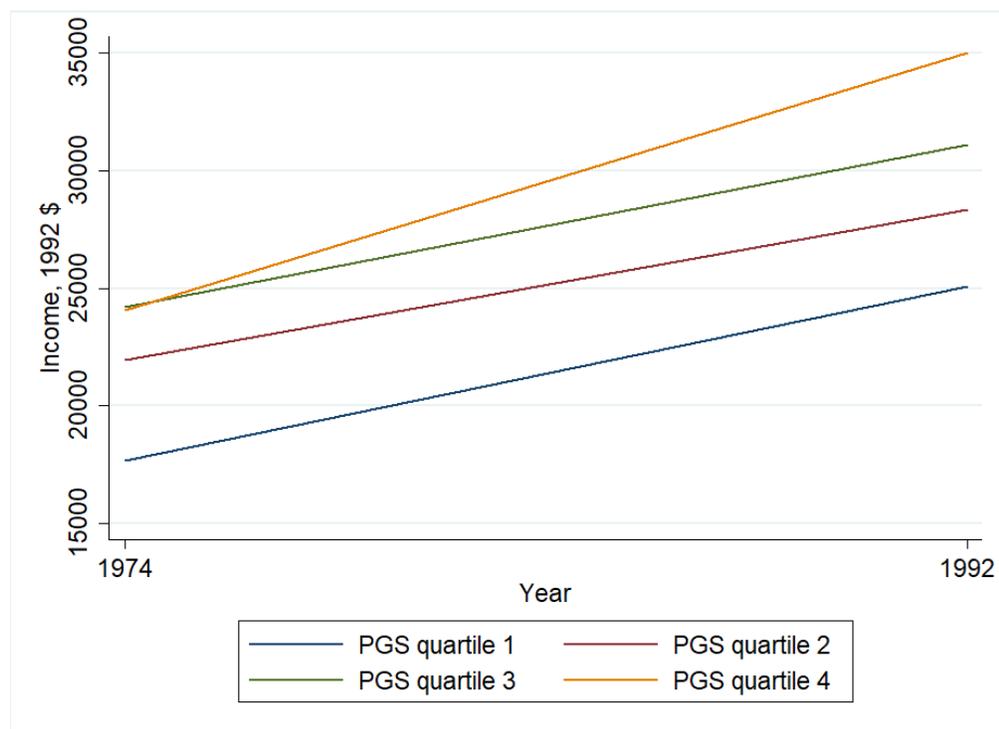

More formally, tables 2-7 present estimates of the effect of a higher polygenic score for educational attainment on mobility in income from wages, annual income from all sources and occupational prestige (Duncan and Siegel scores) in turn. In each case, five specifications are compared. In column (1), a 'control' regression, the dependent variable as measured in 1992 is regressed on the equivalent variable as measured in 1975, age, and gender. Though the PGS is not included as a covariate, the sample is limited to those who consented to provide genetic data. Column (2) adds the polygenic score variable. In column (3), the sample is limited to those with a high school degree only (59% of the sample), and in column (4) the sample is limited to those with more schooling than a high school degree. Column (5) revisits the full sample and includes a fixed effect for the high school variable, meaning estimates are derived from variation within educational category. Regressions in columns (2) through (5) include the first 10 genetic principal components to adjust for population stratification.

Beginning with Table 2, we see in column (2) that a 1-standard deviation in the PGS is associated with nearly $2,000 ($1,986) of additional wage income in 1992, conditional on age, gender and 1974 wage income (measured in 1992 dollars). That is, if we consider two respondents with equal incomes measured in 1974, one of whose PGS is at the 50th percentile and the other of whose PGS is at the 84th percentile, the latter's annual wage income exceeded that of the former by nearly $2,000 in 1992. The estimate is statistically significant at the 1% level. Limiting our results to graduates who did not progress in their education beyond high school in column (3), we find similar results ($1,626). The impact of the PGS in the overall sample is driven by those respondents in the high school only category. When we restrict the sample to those with more than a high school degree in column (4), the coefficient on the PGS variable is statistically indistinguishable from zero. In the education fixed effect model in column (5), the parameter estimate is halved ($887), and is significant only at the 10% level.

Table 2. Wage mobility, 1974-1992, graduate sample

| | (1) Control | (2) Add PGS | (3) High school only | (4) > high school only | (5) Within educ. group |
|---|---|---|---|---|---|
| Birth year | 3341.0*** | 2804.7*** | 996.5 | 2060.2 | 1764.2* |
| | (940.6) | (937.1) | (843.4) | (2052.7) | (930.0) |
| 1974 wages-in $100s | 0.757*** | 0.731*** | 0.503*** | 0.754*** | 0.684*** |
| | (0.0237) | (0.0237) | (0.0293) | (0.0383) | (0.0237) |
| Male | 7020.9*** | 7803.1*** | 7849.1*** | 11494.1*** | 7699.1*** |
| | (1211.9) | (1205.0) | (1202.2) | (2327.8) | (1189.6) |
| PGS | | 1986.0*** | 1626.2*** | -175.0 | 887.3* |
| | | (461.8) | (457.2) | (910.9) | (466.9) |
| Constant | -118102.1*** | -96918.0*** | -26650.2 | -64738.5 | -48965.8 |
| | (36610.5) | (36474.8) | (32812.8) | (80007.6) | (36275.7) |
| Observations | 4546 | 4546 | 2691 | 1855 | 4546 |
| $R^2$ | 0.334 | 0.348 | 0.261 | 0.346 | 0.365 |

Standard errors in parentheses
* p<0.1, ** p<0.05, *** p<0.01

Turning to Table 3, our estimates of the impact of the PGS on total income mobility are still more robust. A 1-standard deviation increase in the PGS is associated with an additional $3,236 increase in 1992. The effect remains and significant in the high school-only sample, and is large, though not statistically significant at conventional levels, in the sample composed exclusively of respondents with more than a high school education. In the education fixed effects model in column (5), we find estimate a highly significant coefficient of considerable magnitude ($1,496).

Table 3. Income mobility, 1974-1992, graduate sample

| | (1)<br>Control | (2)<br>Add PGS | (3)<br>High school only | (4)<br>More than high school | (5)<br>Within educ. group |
|---|---|---|---|---|---|
| Birth year | 4439.6*** | 3714.6*** | 716.5 | 4094.8* | 1967.2* |
| | (1131.7) | (1129.9) | (1026.8) | (2463.1) | (1115.6) |
| 1974 income-in $100s | 0.782*** | 0.756*** | 0.385*** | 0.935*** | 0.715*** |
| | (0.0237) | (0.0237) | (0.0275) | (0.0388) | (0.0235) |
| Male | 6067.5*** | 6957.0*** | 11697.6*** | 3687.5 | 5846.1*** |
| | (1435.7) | (1430.3) | (1420.2) | (2719.4) | (1405.5) |
| PGS | | 3236.3*** | 1674.3*** | 1458.2 | 1496.4*** |
| | | (563.0) | (564.0) | (1092.0) | (566.2) |
| Constant | -155416.7*** | -126950.5*** | -11280.6 | -136697.7 | -47546.5 |
| | (44045.0) | (43977.6) | (39933.1) | (96010.5) | (43517.0) |
| Observations | 5007 | 5007 | 2998 | 2009 | 5007 |
| $R^2$ | 0.298 | 0.310 | 0.194 | 0.342 | 0.336 |

Standard errors in parentheses
* p<0.1, ** p<0.05, *** p<0.01

Results from the model in table 3, column (2) are also presented graphically in Figure 4.  The left-hand panel graphs predicted 1992 incomes by standardized values of the educational attainment PGS, conditional on covariates including 1974 income.  The right-hand panel is a binned scatterplot (Droste, 2019; Stepner, 2013) of the same model.

**Figure 5.** Income (1992) conditional on covariates, by PGS (predictive margins and binned scatterplot).

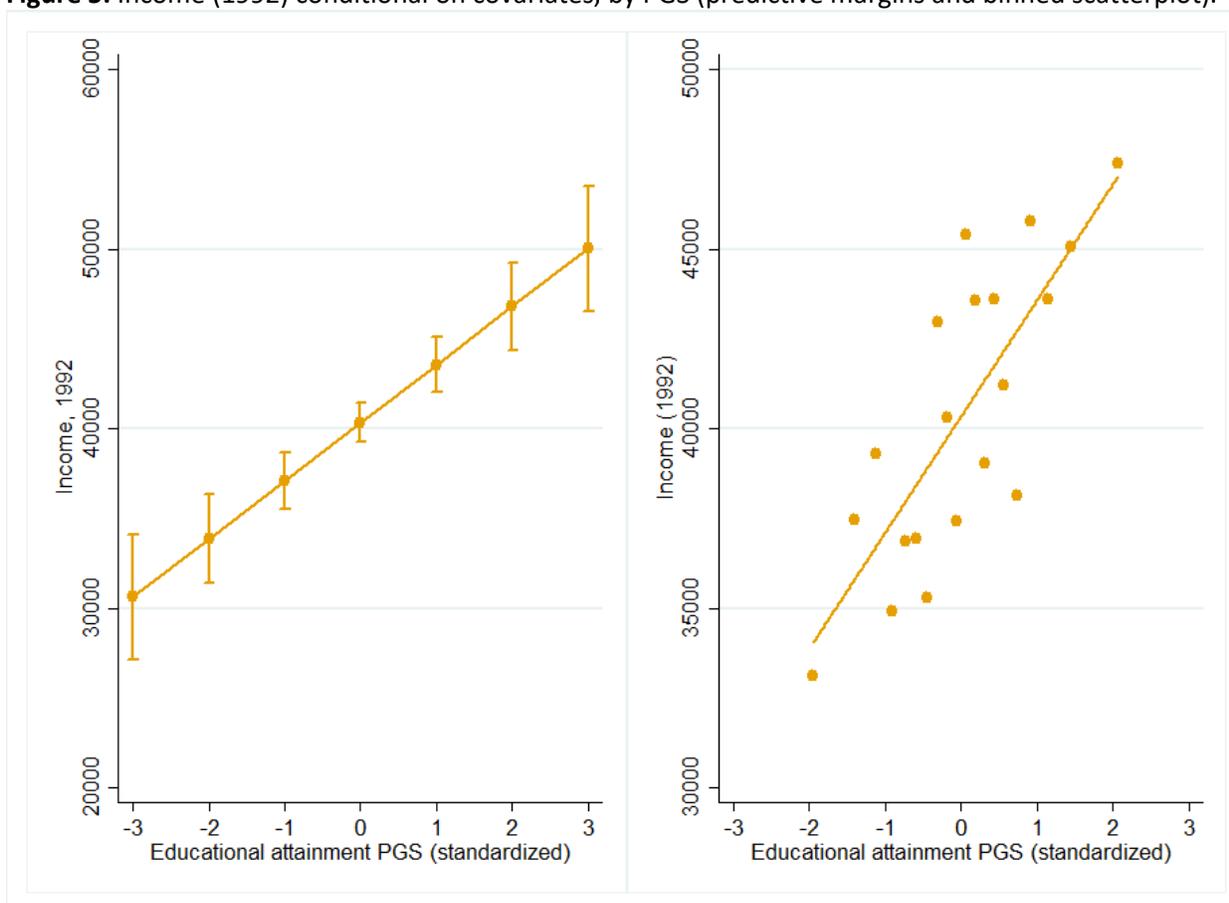

These results show that higher polygenic scores for educational attainment predict wage and income mobility during midlife.  The larger estimated effect of the PGS on total income as compared to wage income is consistent with Barth, Papageorge and Thom's (2020) finding that a higher PGS predicts income in the cross section even after conditioning on wages, suggesting that genes for educational attainment may impact income through facilitating understanding of financial decision-making. Relatedly, Sias, Starks and Turtle (2020) find that a higher PGS predicts stock market participation.

Our estimates in the models including realized education are of particular interest in light of the mechanisms outlined above.  For both wage income and total income, parameter estimates are of larger magnitudes and have lower standard errors in the high school-only sample.  This is consistent with a scenario in which, over time, incomes converge to a long-term mean with a strong genetic component as the impact of noncumulative positive and negative environmental shocks equalizes over time.  We can think of high (low) PGS individuals with low phenotypic education as having experienced disproportionately negative (positive) environmental shocks, which equalize and/or fade out over time, leading to regression to the long-term genetic mean.

In terms of the magnitude of these effects, it should nevertheless be noted that those with only a high school education, whatever their PGS, do not by 1992 attain incomes on par with their peers with more realized education. This is shown in Figure 6, which presents marginal effects after a regression of 1992 income on realized education, the educational attainment PGS, gender and age. For an individual with a PGS three standard deviations above the mean but with only a high school education, the model predicts an income of about $36,000 in 1992. This remains below the predicted income for an individual with an individual scoring three standard deviations below the mean but with more than a high school education (about $50,000). Thus while high PGS individuals had higher incomes and saw higher income mobility, had higher realized education, and within educational categories had higher incomes and greater income mobility, a high PGS does not fully compensate for the lower realized education.

**Figure 6**. Income (1992) by PGS and realized education, marginal effects

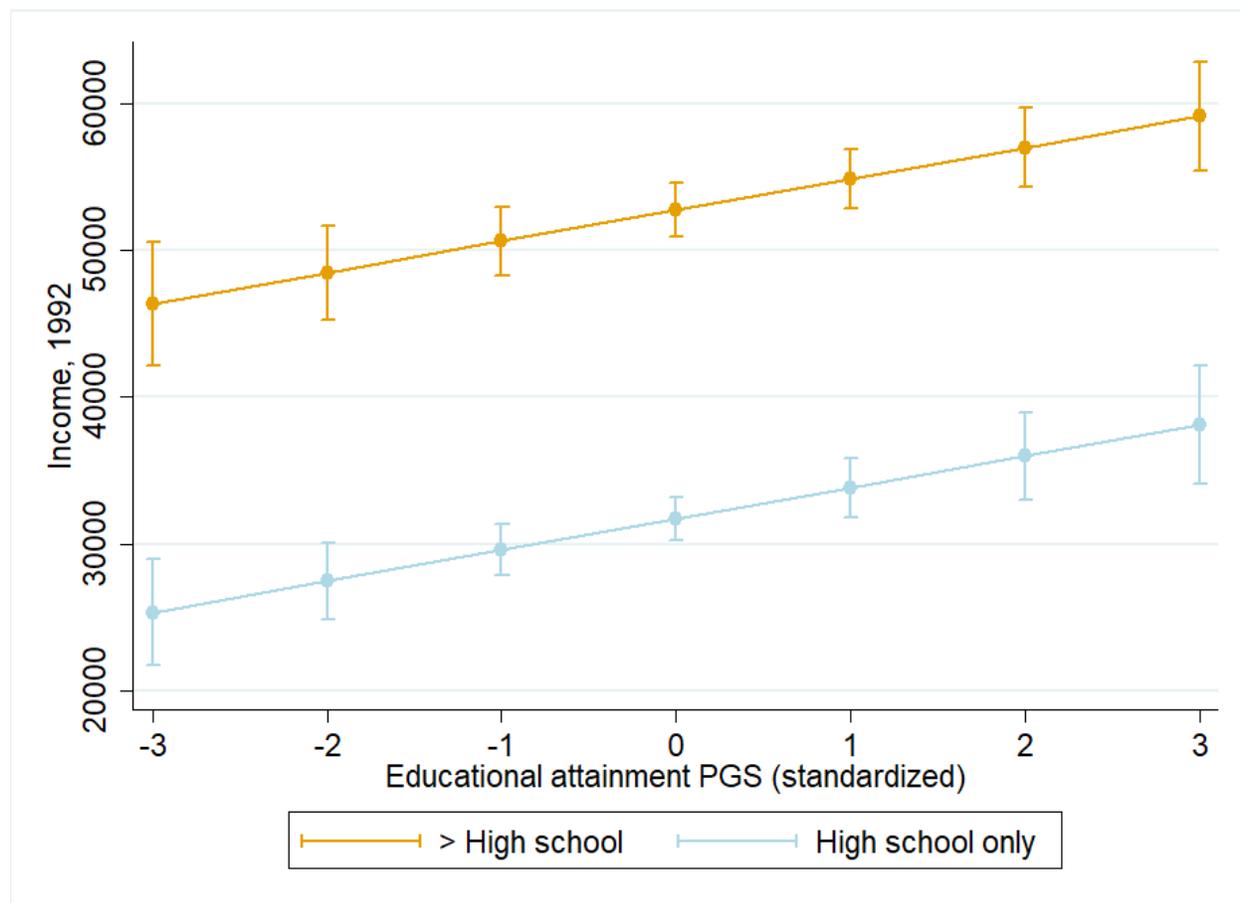

In tables 4 and 5, we turn to mobility in terms of occupational prestige. As predicted, we find that a one standard deviation higher PGS for educational attainment predicts mobility in occupational prestige as measured by both the Duncan (8.4 points, table 4) and Siegel (5.7 points, table 5) prestige scales. To get a sense of the relevant magnitudes, the parameter estimate of the Duncan score is roughly equivalent to moving from employment as a mining engineer to an aerospace engineer, or from a veterinarian to a physician, using 1989 scores as a baseline (Nakao and Treas, 1992). These results show that high-PGS respondents were consistently more likely to move up the occupational prestige ladder. Effects are again concentrated in the high school-only sample.

Table 4. Duncan prestige mobility, 1974-1992, graduate sample

|  | (1) Control | (2) Add PGS | (3) High school only | (4) More than high school | (5) Within educ. group |
|---|---|---|---|---|---|
| Birth year | 21.98*** | 20.60*** | 14.06** | 18.79** | 16.33*** |
|  | (5.227) | (5.241) | (6.484) | (8.429) | (5.125) |
| Duncan score for 1974 job | 0.676*** | 0.667*** | 0.570*** | 0.582*** | 0.581*** |
|  | (0.0114) | (0.0116) | (0.0167) | (0.0199) | (0.0128) |
| Male | -9.001* | -8.528 | -39.63*** | 19.18** | -16.58*** |
|  | (5.259) | (5.262) | (6.922) | (7.780) | (5.167) |
| PGS |  | 8.372*** | 6.682* | -3.349 | 2.469 |
|  |  | (2.685) | (3.676) | (3.770) | (2.652) |
| Constant | -684.8*** | -626.8*** | -347.3 | -480.5 | -362.2* |
|  | (202.8) | (203.4) | (251.8) | (327.6) | (199.4) |
| Observations | 4377 | 4377 | 2569 | 1808 | 4377 |
| $R^2$ | 0.461 | 0.464 | 0.331 | 0.355 | 0.489 |

Standard errors in parentheses
* p<0.1, ** p<0.05, *** p<0.01

Table 5. Siegel prestige mobility, 1974-1992, graduate sample

|  | (1) Control | (2) Add PGS | (3) High school only | (4) More than high school | (5) Within educ. group |
|---|---|---|---|---|---|
| Birth year | 16.11*** | 15.16*** | 9.254** | 17.22*** | 12.06*** |
|  | (3.265) | (3.271) | (3.810) | (5.753) | (3.196) |
| Siegel score for 1974 job | 0.639*** | 0.629*** | 0.505*** | 0.580*** | 0.542*** |
|  | (0.0119) | (0.0121) | (0.0172) | (0.0204) | (0.0132) |
| Male | -2.725 | -2.184 | -18.68*** | 12.13** | -6.557** |
|  | (3.309) | (3.311) | (4.093) | (5.269) | (3.240) |
| PGS |  | 5.694*** | 2.817 | 0.772 | 2.153 |
|  |  | (1.686) | (2.165) | (2.582) | (1.660) |
| Constant | -461.0*** | -419.8*** | -153.1 | -458.5** | -225.1* |
|  | (126.5) | (126.8) | (147.8) | (223.6) | (124.3) |
| Observations | 4377 | 4377 | 2569 | 1808 | 4377 |
| $R^2$ | 0.417 | 0.421 | 0.267 | 0.344 | 0.450 |

Standard errors in parentheses
* p<0.1, ** p<0.05, *** p<0.01

It is interesting to consider whether, conditional on their job when surveyed in 1975, graduates with higher PGSs were more aspirational about the prestige of their future occupations. This question is addressed in tables 6 and 7. In addition to being asked about their current or most recent job, graduate respondents in 1975 were asked to identify the job to which they aspired 10 years' hence. Both the then current and aspired jobs are scored according to their Duncan and Siegel prestige scores. Conditional on age, gender and current job, a one standard deviation increase in PGS is associated with a higher aspired job in both the Duncan (5.4 points) and Siegel (4.3 points) scales, as reported in column (2) of tables 6 and 7, respectively. As seen by comparing columns (3) and (4), this effect is driven by respondents with more than a high school education.

Table 6. Aspired occupational mobility, Duncan score, 1974-1992, graduate sample

|  | (1) Control | (2) Add PGS | (3) High school only | (4) More than high school | (5) Within educ. group |
|---|---|---|---|---|---|
| Duncan score, current job | 0.669*** | 0.648*** | 0.614*** | 0.557*** | 0.595*** |
|  | (0.00862) | (0.0118) | (0.0170) | (0.0215) | (0.0132) |
| Birth year | 11.00*** | 10.04* | 1.393 | 16.79* | 7.072 |
|  | (3.989) | (5.486) | (6.830) | (9.178) | (5.445) |
| Male | -32.79*** | -35.24*** | -44.15*** | -32.75*** | -39.69*** |
|  | (4.007) | (5.426) | (7.201) | (8.338) | (5.400) |
| PGS |  | 5.408** | -2.524 | 6.322 | 1.474 |
|  |  | (2.736) | (3.765) | (4.042) | (2.750) |
| Constant | -215.4 | -165.3 | 171.3 | -350.5 | 9.338 |
|  | (154.7) | (213.0) | (265.4) | (356.7) | (212.0) |
| Observations | 6370 | 3672 | 2094 | 1578 | 3672 |
| $R^2$ | 0.495 | 0.477 | 0.399 | 0.326 | 0.487 |

Standard errors in parentheses
* p<0.1, ** p<0.05, *** p<0.01

Table 7. Aspired occupational mobility, Siegel score, 1974-1992, graduate sample

|  | (1) Control | (2) Add PGS | (3) High school only | (4) More than high school | (5) Within educ. group |
|---|---|---|---|---|---|
| Siegel score, current job | 0.610*** | 0.583*** | 0.526*** | 0.533*** | 0.531*** |
|  | (0.00901) | (0.0121) | (0.0178) | (0.0201) | (0.0133) |
| Birth year | 12.20*** | 12.44*** | 5.322 | 19.44*** | 10.27*** |
|  | (2.463) | (3.319) | (4.017) | (5.770) | (3.292) |
| Male | -14.22*** | -13.90*** | -17.56*** | -16.25*** | -16.51*** |
|  | (2.486) | (3.300) | (4.250) | (5.202) | (3.277) |
| PGS |  | 4.266** | -0.334 | 4.308* | 1.912 |
|  |  | (1.667) | (2.220) | (2.554) | (1.669) |
| Constant | -257.7*** | -253.0** | 36.31 | -483.9** | -123.9 |
|  | (95.37) | (128.7) | (156.0) | (224.2) | (128.1) |
| Observations | 6370 | 3672 | 2094 | 1578 | 3672 |
| $R^2$ | 0.432 | 0.424 | 0.305 | 0.341 | 0.437 |

Standard errors in parentheses

* p<0.1, ** p<0.05, *** p<0.01

Our results to this point are consistent with a "wash-out effect" whereby shocks from the non-shared environment are of temporary and non-cumulative importance, or with self-directed nurture of nature effects (Plomin et al., 1977). They do not preclude a scenario where early parental investments, a nurture of nature effect, slowly mature and predict continued economic mobility during prime earning years. Papageorge and Thom (2020) find that the educational attainment PGS more strongly predicts college completion among children growing up in higher socioeconomic status households, and Muslimova et al. (2020) show that the PGS is more predictive of years of education among first born children, who are thought to receive greater parental investments. In each case, parental investments, which presumably increase with family socioeconomic status, interact with genetic predispositions via a nurture of nature effect. In the WLS data, although it is a priori unlikely that these investments had not matured by the time respondents were approximately 36 years old, it is worthwhile to test for interactions between respondents' PGS and family socioeconomic status in determining economic outcomes. If, even at the mid-career stage, increasing mobility of high PGS respondents is driven by early parental investments, a nurturing of nature, then the coefficient on an interaction between parental socioeconomic status and the PGS variable in the mobility regressions above should be positive.

In the 1975 survey wave, graduate respondents were asked questions relating to parents' social status in 1957, the year the graduates finished high school. I separately examine interactions between the PGS and four measures of family socioeconomic status: father's years of education, mother's years of education, father's Duncan occupational prestige score, mother's Duncan occupational prestige score. Cases where parents did not work in 1957 are dropped from the latter two regressions. As dependent variables, I consider the same four measures of economic status as above: wages, income, Duncan occupational prestige and Siegel occupational prestige in 1992, conditional on their 1974 values and the gender and age covariates. Results are presented in tables 8-11. I find statistically insignificant interactions between the PGS variable and father's education across all mobility measures (table 8) and mother's education (table 9). Indeed, the parameter estimates are as often negative as positive. The interaction between the PGS and father's educational prestige is in fact negative in terms of father's occupational prestige (table 10) and mother's occupational prestige (table 11). No significant interactions are observed then the dependent variables are wages or income. This absence interactions between parental social status and the PGS in shaping mid-life outcomes is an interesting contrast to the early life results found by Muslimova et al. (2020) and Papageorge and Thom (2020).

Table 8. Economic mobility, 1975-1992, interaction with father's education

| | (1) Wages | (2) Income | (3) Duncan | (4) Siegel |
|---|---|---|---|---|
| Birth year | 2536.5*** | 3160.6*** | 20.98*** | 14.55*** |
| | (963.6) | (1159.4) | (5.363) | (3.342) |
| 1974 wages | 0.727*** | | | |
| | (0.0242) | | | |
| Male | 7975.7*** | 7366.9*** | -7.053 | -1.878 |
| | (1232.0) | (1454.6) | (5.341) | (3.351) |
| PGS | 1929.6 | 2123.7 | 14.00* | 9.317* |
| | (1391.9) | (1697.9) | (8.098) | (5.054) |
| Father education (years) | 572.4*** | 985.7*** | 4.647*** | 3.077*** |
| | (140.6) | (170.6) | (0.806) | (0.500) |
| Father education*PGS | -23.75 | 70.71 | -0.711 | -0.431 |
| | (130.8) | (159.3) | (0.757) | (0.472) |
| 1974 Income | | 0.749*** | | |
| | | (0.0240) | | |
| 1974 Duncan prestige | | | 0.651*** | |
| | | | (0.0121) | |
| 1974 Siegel prestige | | | | 0.615*** |
| | | | | (0.0125) |
| Constant | -91896.4** | -114802.9** | -677.9*** | -419.2*** |
| | (37441.9) | (45031.6) | (207.9) | (129.4) |
| Observations | 4418 | 4841 | 4226 | 4226 |
| $R^2$ | 0.350 | 0.317 | 0.465 | 0.425 |

Standard errors in parentheses
* p<0.1, ** p<0.05, *** p<0.01

Table 9. Economic mobility, 1975-1992, interaction with mother's education

| | (1) Wages | (2) Income | (3) Duncan | (4) Siegel |
|---|---|---|---|---|
| Birth year | 2334.6** | 2933.0** | 17.92*** | 13.19*** |
| | (947.4) | (1142.5) | (5.288) | (3.300) |
| 1974 wages | 0.731*** | | | |
| | (0.0238) | | | |
| Male | 7721.4*** | 6963.1*** | -8.794* | -2.079 |
| | (1210.0) | (1435.3) | (5.280) | (3.319) |
| PGS | 5114.4*** | 2903.3 | 13.26 | 9.921 |
| | (1722.4) | (2108.3) | (10.19) | (6.363) |
| Mother education (years) | 653.1*** | 1193.0*** | 5.132*** | 3.313*** |
| | (170.0) | (207.5) | (0.981) | (0.612) |
| Mother education*PGS | -309.1** | 1.660 | -0.500 | -0.411 |
| | (155.7) | (190.7) | (0.916) | (0.572) |
| 1974 Income | | 0.752*** | | |
| | | (0.0238) | | |

| | (1) Wages | (2) Income | (3) Duncan | (4) Siegel |
|---|---|---|---|---|
| 1974 Duncan prestige | | | 0.657*** | |
| | | | (0.0118) | |
| 1974 Siegel prestige | | | | 0.619*** |
| | | | | (0.0123) |
| Constant | -85265.1** | -108882.5** | -571.4*** | -373.7*** |
| | (36740.3) | (44283.7) | (204.5) | (127.5) |
| Observations | 4512 | 4968 | 4338 | 4338 |
| $R^2$ | 0.352 | 0.315 | 0.467 | 0.425 |

Standard errors in parentheses
* p<0.1, ** p<0.05, *** p<0.01

## Table 10. Economic mobility, 1975-1992, interaction with father's occupational prestige

| | (1) Wages | (2) Income | (3) Duncan | (4) Siegel |
|---|---|---|---|---|
| Birth year | 2458.8*** | 3190.5*** | 20.10*** | 14.81*** |
| | (940.8) | (1133.3) | (5.257) | (3.294) |
| 1974 wages | 0.708*** | | | |
| | (0.0238) | | | |
| Male | 8652.4*** | 7804.5*** | -8.300 | -2.118 |
| | (1209.3) | (1429.2) | (5.253) | (3.316) |
| PGS | 1503.0* | 2252.5** | 18.90*** | 10.21*** |
| | (833.5) | (1013.6) | (4.774) | (2.997) |
| Father education (years) | 14.98*** | 23.61*** | 0.0934*** | 0.0425*** |
| | (2.159) | (2.615) | (0.0126) | (0.00783) |
| Father's occupational prestige*PGS | 0.840 | 1.936 | -0.0338*** | -0.0146** |
| | (2.029) | (2.475) | (0.0117) | (0.00734) |
| 1974 Income | | 0.736*** | | |
| | | (0.0237) | | |
| 1974 Duncan prestige | | | 0.645*** | |
| | | | (0.0120) | |
| 1974 Siegel prestige | | | | 0.614*** |
| | | | | (0.0125) |
| Constant | -88273.2** | -114249.8*** | -626.5*** | -413.1*** |
| | (36592.2) | (44071.7) | (204.0) | (127.7) |
| Observations | 4511 | 4969 | 4341 | 4341 |
| $R^2$ | 0.356 | 0.321 | 0.470 | 0.424 |

Standard errors in parentheses
* p<0.1, ** p<0.05, *** p<0.01

## Table 11. Economic mobility, 1975-1992, interaction with mother's occupational prestige

| | (1) Wages | (2) Income | (3) Duncan | (4) Siegel |
|---|---|---|---|---|
| Birth year | 1127.3 | -665.2 | 13.49* | 10.92** |
| | (1414.2) | (1742.6) | (8.078) | (5.048) |
| 1974 wages | 0.708*** | | | |
| | (0.0352) | | | |
| Male | 5194.0*** | 6201.9*** | -8.569 | -1.867 |

| | (1) | (2) | (3) | (4) |
|---|---|---|---|---|
| | (1879.2) | (2328.4) | (8.494) | (5.314) |
| PGS | 2123.6 | 4909.0*** | 20.54** | 15.44*** |
| | (1389.0) | (1733.4) | (8.256) | (5.174) |
| Father education (years) | 13.55*** | 15.30*** | 0.0885*** | 0.0464*** |
| | (3.583) | (4.492) | (0.0213) | (0.0133) |
| Mother's occupational prestige*PGS | -2.896 | -5.438 | -0.0365* | -0.0241* |
| | (3.361) | (4.244) | (0.0204) | (0.0128) |
| 1974 Income | | 0.746*** | | |
| | | (0.0398) | | |
| 1974 Duncan prestige | | | 0.637*** | |
| | | | (0.0191) | |
| 1974 Siegel prestige | | | | 0.603*** |
| | | | | (0.0195) |
| Constant | -35395.0 | 38505.5 | -363.7 | -258.0 |
| | (54951.2) | (67672.7) | (313.1) | (195.5) |
| Observations | 1647 | 1823 | 1603 | 1603 |
| $R^2$ | 0.374 | 0.323 | 0.467 | 0.435 |

Standard errors in parentheses
* p<0.1, ** p<0.05, *** p<0.01

### 3.2.2. Sibling-pair sample

As noted above, a positive association between a PGS for educational attainment and an economic phenotype of interest in a sample of unrelated individuals could be spurious. As parents may transmit financial and cultural capital to offspring along with their genes, a PGS-phenotype association could be confounded with inherited social class advantage. This concern is mitigated in the above analysis as all models include the first 10 genetic principal components. Moreover, it should be emphasized that the DNA samples used to construct the PGS index in Lee et al. (2018) were all drawn decades after the measurement of economic variables in the WLS data, and the genetic data of WLS respondents was not used in the GWAS from which the genetic weights are derived. Thus, the above analysis is entirely out of sample.

Nevertheless, a causal interpretation of the coefficient on the educational attainment PGS is more defensible in the context of a sibling-pair design, which controls for family-level fixed effects experienced equally by both siblings. As noted above, the WLS surveyed a randomly selected sibling of graduate respondents in the 1977 survey wave, and these siblings were surveyed again in 1993-94. Though the data collected from siblings is less comprehensive than that collected from the graduate sample, we do have data on the sibling's total annual income in these two survey waves, which followed the corresponding wave of graduate surveys by 1-2 years. Results for the sibling-pair analysis are presented in Table 12. Column (1) reproduces the previous graduate-only analysis in this new sample, pooling all siblings and graduates together. The coefficient on PGS ($2599) is, as expected, positive and statistically significant at p=0.01. The remaining columns present estimates from family fixed effects models. In column (2), the PGS variable represents a simple rank, wherein the sibling with the lower PGS is assigned a value of 1, and the sibling with the higher rank is assigned a 2. In column (3), the PGS variable represents each respondent's standardized PGS (standardized within the pooled graduate-sibling sample). We estimate effects of similar magnitudes ($3,016 - $3,619) in this sibling-pair analysis as in the income mobility results reported in Table 3 for the graduate-only sample. The estimate is significant at p=0.1. All models include controls for age and gender. In columns (4) and (5), the income variables are log-transformed after adding 1 to the original value. The PGS operationalizations are the within-sibling rank and normalized score in columns (4) and (5), respectively. The PGS estimate remains positive as expected, and statistically significant at p=0.01 and p=0.1, respectively. In sum, given our controls for age and gender, we observe that even controlling for family fixed effects, among two same-sex siblings of the same age, the sibling with the higher PGS for educational attainment experienced more income mobility between the mid-1970s and early 1990s, conditional on income in the earlier period.

Table 12. Income mobility, 1974-1992, sibling-pair sample

|  | (1) Pooled OLS | (2) Levels, PGS rank | (3) Levels, PGS score | (4) Logs+1, PGS rank | (5) Logs+1, PGS score | (6) Levels, interaction | (7) Logs+1, interaction |
|---|---|---|---|---|---|---|---|
| Income, 1974, ($100s). | 0.575*** | 0.365*** | 0.363*** |  |  | 0.367*** |  |
|  | (0.0252) | (0.0988) | (0.0992) |  |  | (0.0971) |  |
| Age in 1975/7 survey round | -1340.7*** | -1034.6*** | -1009.6*** | -0.230*** | -0.227*** | -1012.9*** | -0.231*** |
|  | (198.2) | (253.7) | (253.2) | (0.0345) | (0.0348) | (260.8) | (0.0347) |
| Male | 9541.2*** | 17741.0*** | 17633.2*** | 0.705* | 0.637 | 18055.2*** | 0.741* |
|  | (1560.4) | (4149.6) | (4148.9) | (0.409) | (0.411) | (4180.6) | (0.410) |
| PGS | 2599.3*** |  | 3016.4* |  | 0.388* | -5040.2 | -0.495 |
|  | (635.0) |  | (1773.9) |  | (0.223) | (4371.6) | (0.479) |
| PGS rank within sib. pair |  | 3619.9* |  | 0.569*** |  | 3734.3 | 0.783** |
|  |  | (1916.0) |  | (0.207) |  | (3286.8) | (0.336) |
| (Log) 1974 income |  |  |  |  |  |  |  |
| (Log+1) 1974 income |  |  |  | 0.368*** | 0.382*** |  | 0.363*** |
|  |  |  |  | (0.0991) | (0.0989) |  | (0.0995) |
| PGS*PGS rank |  |  |  |  |  | 3074.5 | 0.133 |
|  |  |  |  |  |  | (1982.2) | (0.229) |

| Constant | 60930.1*** | 46684.5*** | 51176.8*** | 14.54*** | 15.25*** | 44975.8*** | 14.23*** |
|---|---|---|---|---|---|---|---|
| | (7084.8) | (9450.4) | (9159.8) | (1.264) | (1.250) | (10384.9) | (1.309) |
| Observations | 2492 | 2492 | 2492 | 2492 | 2492 | 2492 | 2492 |
| $R^2$ | 0.352 | 0.216 | 0.214 | 0.174 | 0.167 | 0.219 | 0.175 |

Standard errors in parentheses.  Standard errors clustered at sibling family level in columns (2) through (7).
* p<0.1, ** p<0.05, *** p<0.01

Existing studies linking educational attainment PGSs with phenotypes of interest have tended to find that coefficients on the PGS fall significantly in sibling-pair designs relative to samples of non-relatives. As a predictor of phenotypic educational attainment, Domingue et al. (2015) and Fletcher et al. (2020) estimate effect sizes among siblings about half that found between non-relatives.  Similar attenuations are estimated by Belsky et al. (2018) in their study of educational mobility relative to parents.  The findings of our within-sibling analysis here show that the magnitude of the impact of genes associated with educational attainment on income mobility within middle age is robust to controls for the shared childhood environment.

A notable finding in Fletcher et al. (2020) is that parents engage in compensatory investment in an attempt to equalize outcomes among their children.  That is, parents intuit ability differences among their children, and invest a higher proportion of scarce resources in the lower ability child.  This behaviour is a potentially important source of nonshared environmental differences among siblings.  In their sibling-pair analysis of educational attainment, Fletcher et al. (2020) provide evidence for a compensatory effect by interacting the PGS variable with a variable indicating the rank of an individual's PGS within their sibling pair.  Consistent with a compensatory effect, they estimate a negative coefficient on this interaction term, indicating that the PGS is less predictive of educational attainment for the higher-ranked sibling.

The compensatory effect is an instance of a predictable non-shared environmental effect on educational attainment.  The accounts of increasing trait heritability with age discussed above predict that nonshared environmental effects either fade out with time or are uncorrelated and increasingly mean-zero with time.  By the time children are in the middle of their adult working lives, these accounts predict that compensatory investments made by parents in childhood and young adulthood have depreciated in value.  Drawing on Fletcher et al. (2020), columns (6) and (7) of Table 8 present estimates from models where each sibling's rank within their sibling pair is interacted with their educational attainment PGS, with the dependent variable measured in levels and logs, respectively.  The coefficient on the interaction term is not statistically significant, and indeed the parameter estimates are positive. Whatever compensatory investments the parents of our sample have made during their children's formative years, these do not appear to attenuate the effect of sibling genetic differences as predictors of income mobility within adulthood.

### 3.3. Extensions

#### 3.3.1. Mobility mechanisms

It is interesting to consider the mechanisms underlying the results presented above. Conceptually, the educational attainment PGS captures a tendency to acquire human capital. This could be reflected in movement between firms or industries, or in changing job duties within firms. In keeping with Sias, Starks and Turtle's (2020) results on stock market participation and financial planning, individuals with higher PGSs may also be more forward looking, leading to higher levels of human capital investment at each stage of their career. They might also be better at forecasting which industries are likely to grow in the future, and shifting into these industries over their career.

##### 3.3.1.1. Firm and industry change

I first consider whether higher scoring individuals were more likely to change industries or employers between 1974 and 1992. The WLS data enable us to identify whether graduate respondents changed employers across these years and the number of employers across these years. In table 13, I divide the sample by PGS quartile, and compare the frequency with which respondents had one employer throughout the entire period, as opposed to more than one employer. Higher PGS quartiles do not predict having had more than one employer.

Table 13. Frequency table of more than one employer, by PGS quartile

| PGS quartile | One employer | >1 employer | |
|---|---|---|---|
| 1 | 44.38 | 55.62 | 100.00 |
| 2 | 44.17 | 55.83 | 100.00 |
| 3 | 44.77 | 55.23 | 100.00 |
| 4 | 44.09 | 55.91 | 100.00 |
| Total | 44.35 | 55.65 | 100.00 |
| $N$ | 5048 | | |

More formally, table 14 presents the results of logistic regressions of having more than one employer on covariates, including the PGS. Column (1) does not include controls for having attained more than a high school education, whereas column (2) includes fixed effects for realized education. Columns (3) and (4) are ordered logistic regressions with number of employers as the dependent variable. All coefficients are reported as odds ratios. The PGS is not predictive of the number of employers a respondent reports between 1974 and 1992.

Table 14. Logistic and ordered logit regressions, number of employers

| | (1) Logit, >1 employer | (2) Logit, >1 employer | (3) Ordered logit, no. employers | (4) Ordered logit, no. employers |
|---|---|---|---|---|
| Male | 0.587*** | 0.579*** | 0.611*** | 0.598*** |
| | (-9.24) | (-9.35) | (-9.35) | (-9.64) |
| Age | 1.168** | 1.157* | 1.140* | 1.122* |
| | (2.67) | (2.49) | (2.46) | (2.15) |
| PGS | 0.979 | 0.969 | 0.989 | 0.972 |
| | (-0.74) | (-1.06) | (-0.44) | (-1.06) |
| HS FEs | No | Yes | No | Yes |
| $N$ | 5048 | 5048 | 5048 | 5048 |

Exponentiated coefficients; $t$ statistics in parentheses
$^{*} p < 0.05$, $^{**} p < 0.01$, $^{***} p < 0.001$

There is some evidence, however, that high PGS respondents with only a high school degree were more likely to remain with an employer in the early stages of their career. 1975 respondents were asked the number of months they had been with their current employer. In column (1) of table 15, we see that in the full sample of graduate respondents, the PGS is not predictive of increased tenure with current employer. In column (2), the sample is limited to respondents with only a high school degree. In column (3), the sample is limited to those with more than a high school degree. We can see that among respondents with only a high school degree, a higher PGS is predictive of about an additional four months spent with the current employer.

Table 15. Length of tenure with 1975 employer, regression results.

|  | (1) Full sample | (2) HS only | (3) >HS |
|---|---|---|---|
| Male | 43.82*** | 57.50*** | 31.69*** |
|  | (25.55) | (24.82) | (12.97) |
| Age | -8.089*** | -7.765*** | 1.305 |
|  | (-4.70) | (-3.58) | (0.48) |
| PGS | -1.019 | 3.788** | -1.345 |
|  | (-1.19) | (3.10) | (-1.15) |
| _cons | 357.6*** | 347.4*** | -11.92 |
|  | (5.34) | (4.11) | (-0.11) |
| *N* | 4392 | 2605 | 1787 |

*t* statistics in parentheses
* $p < 0.05$, ** $p < 0.01$, *** $p < 0.001$

High PGS respondents were also no more likely to change industries, as shown in Table 16. The dependent variable for this logistic regression is a dummy taking the value of 1 if the respondent ever changed industries across their employers between 1974 and 1992. Column (1) pools graduate respondents together, and column (2) includes a fixed effect for having completed only a high school education. The PGS variable is not predictive of changing industries during a respondent's career.

Table 16. Logistic regressions, ever changed industry

|  | (1) Ever changed ind., pooled | (2) Ever changed ind., H.S. only FEs |
|---|---|---|
| PGS | 0.978 | 0.995 |
|  | (-0.80) | (-0.17) |
| Male | 0.581*** | 0.594*** |
|  | (-9.55) | (-9.04) |
| Age | 1.117 | 1.135* |
|  | (1.92) | (2.18) |
| *N* | 5387 | 5387 |

Exponentiated coefficients; *t* statistics in parentheses
* $p < 0.05$, ** $p < 0.01$, *** $p < 0.001$

Though high PGS respondents were not more likely to switch industries, it could be the case that they were more likely to begin their careers in industries that would subsequently see high wage growth. I approach this question using US census data on income and occupation via IPUMS (Ruggles et al., 2022)[6]. Using 1990 industry codes, I calculate median wages both in 1990 and 1980 for each industry. I then compute the change in median incomes for each industry, and rank industries according to this change. After merging with the WLS data, we have, for each respondent, data on the industry in which they were employed in 1974, a variable whose value indicates the rank of that industry in terms of median wage growth from 1980 to 1990, and a variable which is the difference in median wages in the industry from 1980 to 1990. For the ranking variable, higher values indicate greater median wage growth. An extract of the resulting data will perhaps clarify the structure of this data (Table 17).

Table 17. Data extract indicating structure after merging with IPUMS data

| Id (WLS) | PGS (WLS) | Industry (WLS; IPUMS) | Median income change in industry, '80-'90 (IPUMS) | Rank of median income change among industries, '80-'90 (IPUMS) |
|----------|-----------|------------------------|---------------------------------------------------|----------------------------------------------------------------|
| 902029 | 0.47273 | 100 | 3995 | 50 |
| 914884 | -0.00564 | 111 | 5995 | 102 |
| 914724 | 2.472495 | 121 | 4495 | 64 |
| 926685 | 0.000796 | 122 | 7995 | 153 |
| 900488 | 0.455838 | 130 | 16495 | 217 |

In separate regressions, I then regress each of these variables on the educational attainment PGS and the usual controls for age and gender. Results of this exercise using the income variable are presented via binned scatterplots in figure 7. The left-hand panel does not include a control for realized education, whereas the panel on the right does. The figure shows that high PGS respondents were more likely, in 1975, to be working in industries which subsequently experienced high wage growth. Put differently, the binned scatterplot sorts respondents into bins according to the PGS variable, and then calculates the mean median wage changes across the industries in which respondents from this bin worked. Though full tabular results omitted for space considerations, I find that a 1-standard deviation increase in the educational attainment PGS predicts being, in 1974, in an industry which would see an average increase of $121 in median income from 1980 to 1990, relative to a lower-scoring PGS respondent.

---

**Figure 7**. Binned scatterplots, mean median income change within industry by PGS

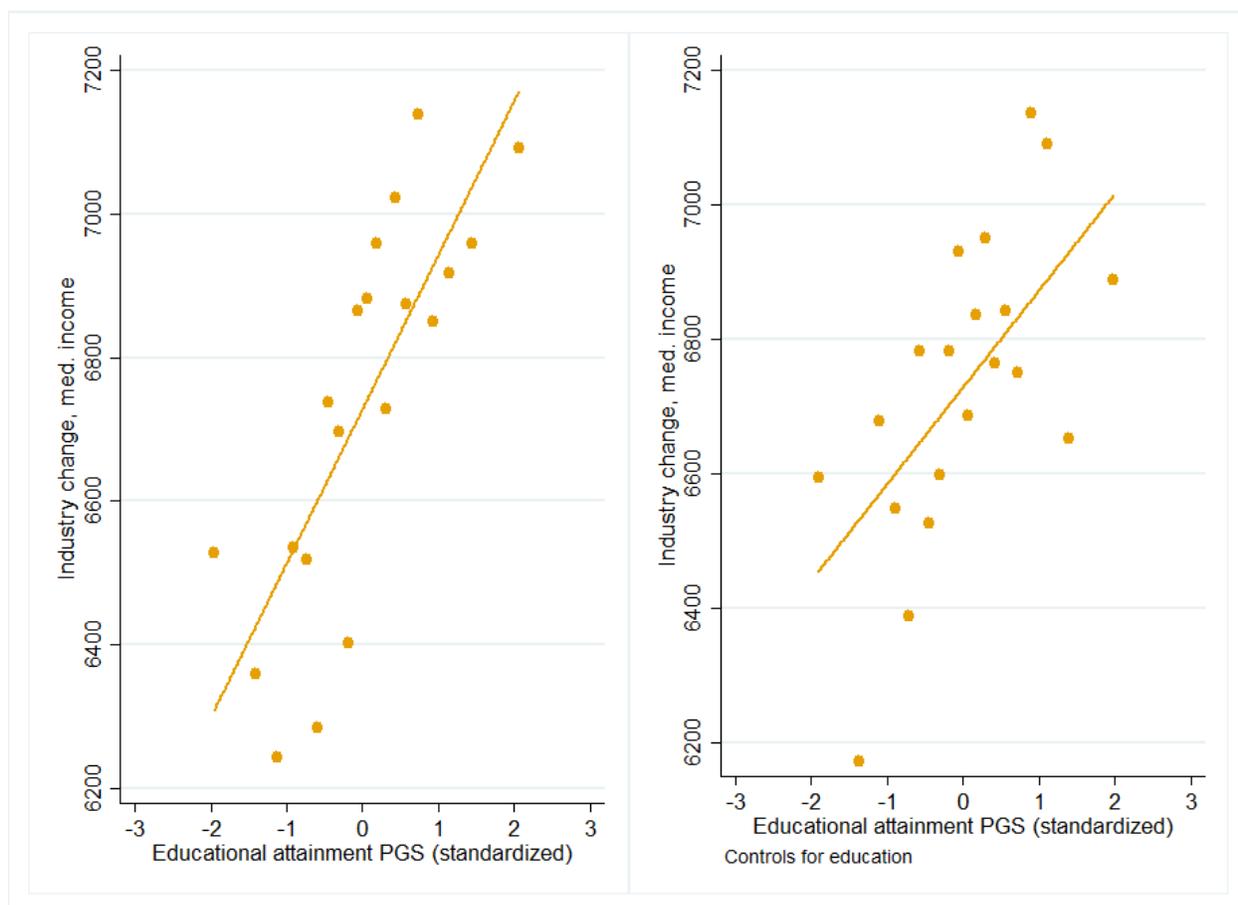

### 3.3.1.2. Human capital accumulation with employers

I next consider whether higher PGS respondents were more likely to accumulate human capital during their career. In the 1992 survey wave, graduates were asked a series of questions about their job spells with up to four employers. Of interest are questions about whether the respondent took, "any training or classes that could help him/her to find another job", and on whether the respondent had a change in, "his/her most important job activities or duties". I first estimate a logistic regression in which the dependent variable is a dummy indicating having taken training during the respondent's tenure with their first employer. This exercise is then repeated for training during their tenure with their second employer, and so on. Although not all respondents had an equal number of employers, and although one person's first employer spell could overlap in time with, for instance, another respondent's second employer spell, we will recall that the PGS is not associated with number of employers between 1974 and 1992.

Parameter estimates for the PGS variable are presented graphically in the left-hand panel of figure 8 in the form of odds ratios. Unsurprisingly, since fewer respondents had four employers than one, estimates are less precise as we move from the first to subsequent employers. Nevertheless, we estimate consistently positive coefficients on the PGS variable, in the range of a 5-10% higher likelihood of taking training during tenure with an employer. The right-hand panel reports estimates from models in which the dependent variable is a dummy indicating that the respondent changed their main duties during each of their employment spells. Here results are more robust, as the PGS predicts a greater likelihood of changing main job duties. All models include gender and age controls.

**Figure 8**. Coefficient plots, likelihood of changing job duties (odds ratios), full sample

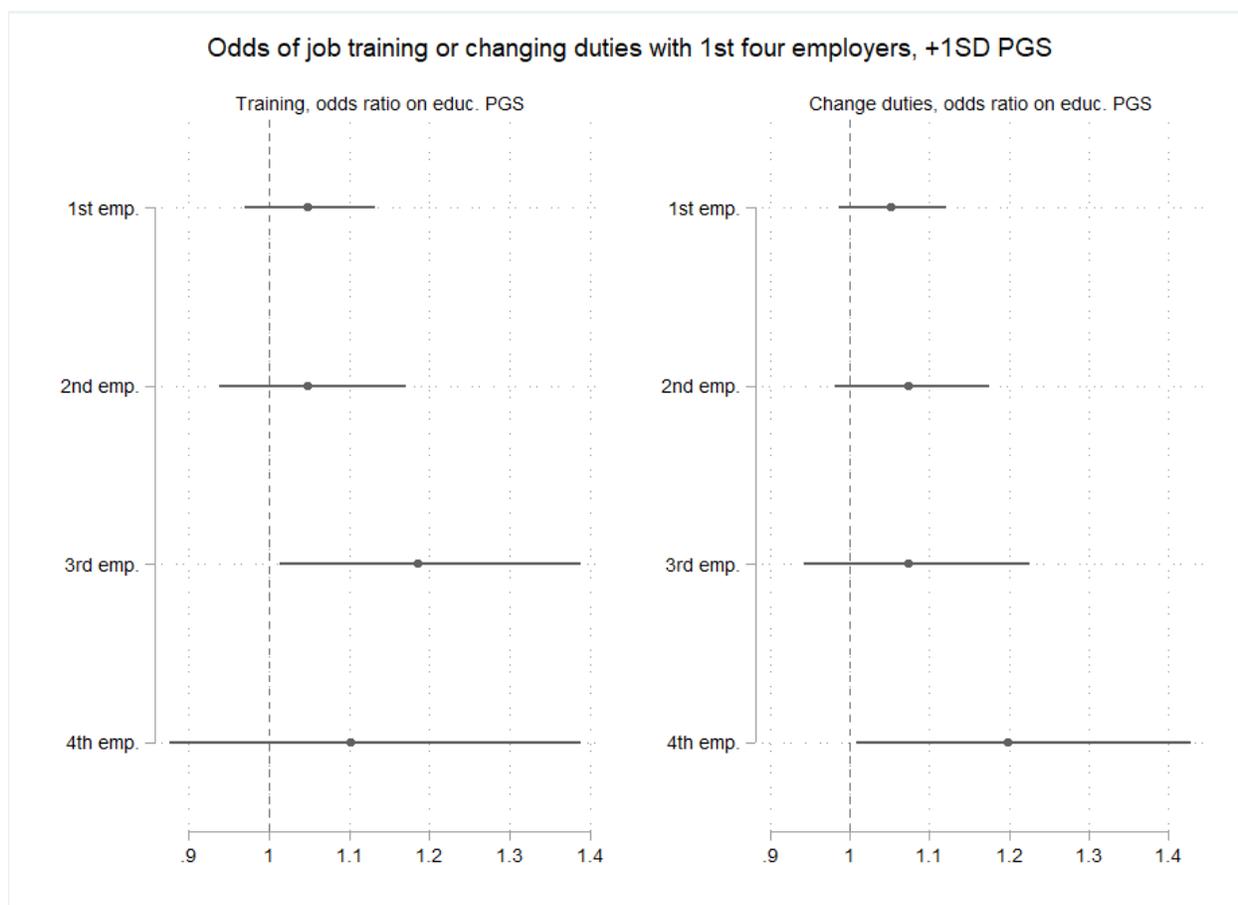

Given the variation in our main results by phenotypic education, I replicate this analysis in a sample comprised only of those with no more than a high school education. Estimates are presented in figure 9. On the whole, results are more robust, indicating that high PGS individuals with lower realized education were particularly likely to take further training and to change duties during a job spell with an employer. Though imprecisely estimated, our estimates within this sample are in the range of 10-25% higher likelihood of training or changing job duties.

**Figure 9**. Coefficient plots, likelihood of changing job duties (odds ratios), high school only sample

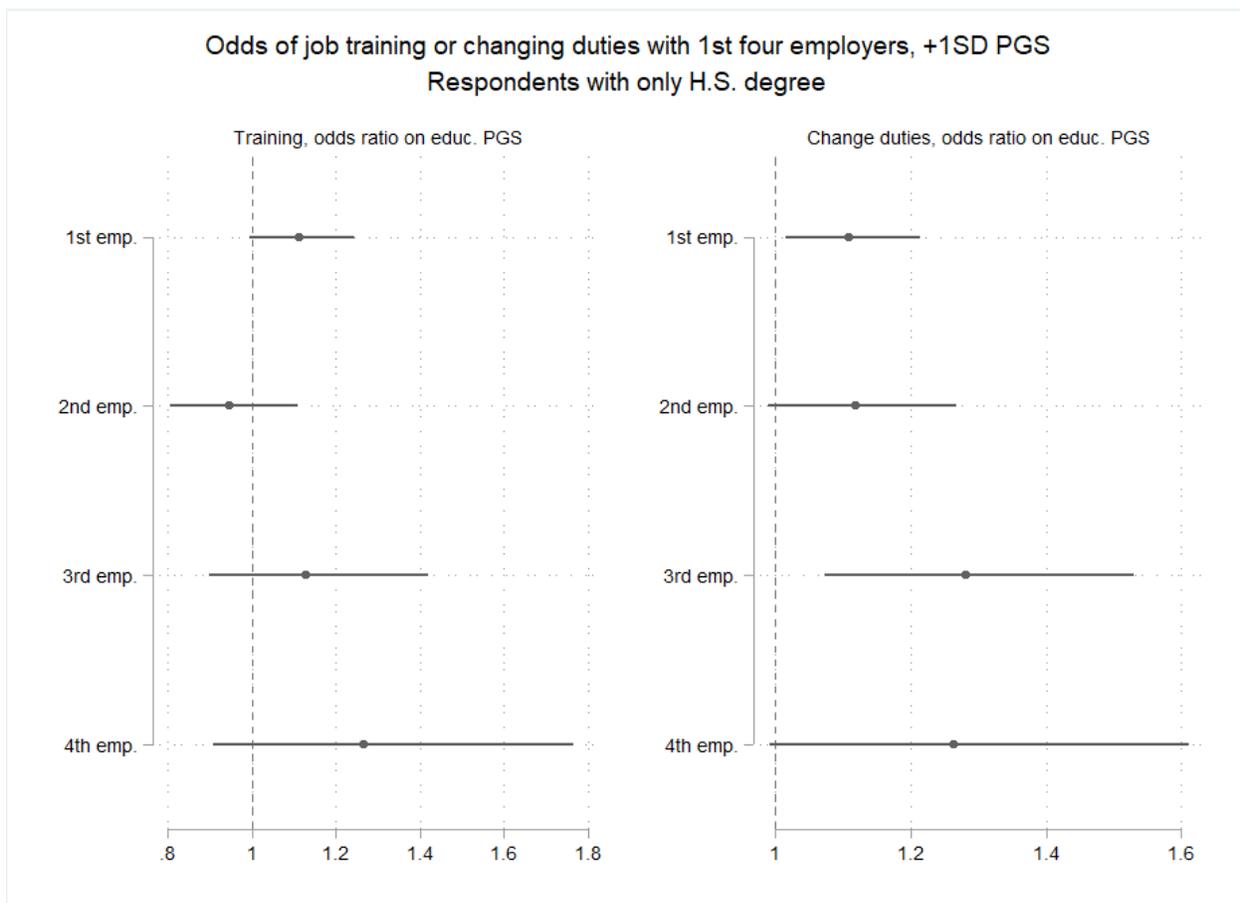

4. Discussion

Leveraging insights into the genetic architecture of complex behavioural phenotypes drawn from large datasets, a recent body of literature has found, in support of existing twin studies, that many economic outcomes have a significant genetic component. Previous research has shown that a polygenic score for educational attainment predicts income, wealth, and stock market participation (Barth, Papageorge and Thom, 2020; Sias, Starks and Turtle 2020), and social mobility across generations (Belsky et al. 2018) in samples not used to construct the PGSs. As estimates are based on direct reading of human genomes, their causal interpretation does not rest on the assumptions of traditional twin studies. The latter result, being estimated via a sibling-pair analysis, still more plausibly has a causal interpretation. In this paper, I show that an educational attainment PGS predicts wage, income and occupational prestige mobility across the prime working years among 1957 Wisconsin high school graduates, conditional on early career status. Moreover, in a sibling-pair design, the sibling with the higher PGS tended to experience more income mobility than the sibling with the lower PGS. High-PGS graduates were also more aspirational about their occupational mobility ten years hence when surveyed in 1975. I find no evidence that parental socioeconomic status mediates the relationship between the PGS and mid-career mobility. Nor do I find evidence, in the sibling-pair design, that the PGS effect is mediated by a sibling's rank within the sibling pair. These latter two findings are evidence against a nurture of nature effect driving our results. I argue they are more consistent with direct genetic effects, and a scenario where the effect of nonshared environmental shocks is non-cumulative and "wash-out" over time. The results are also consistent with a scenario in which individuals, with age, are better able to structure their environments in a manner complementary to their genotype, a form of what Plomin et al. (1977) referred to as "active correlation" between genotypes and environments.

These results speak to observed heterogeneity in workers' responses to skill-biased technological change, particularly given the time frame of the WLS data. I find that high PGS individuals were more likely, early in their careers, to be working in industries which would later be characterized by high median wage growth. Consistent with earlier research (e.g., Barth, Papageorge and Thom, 2020, Sias, Starks and Turtle 2020), high PGS individuals may be more forward looking and more comfortable with forecasting returns to acquired skills. As expected, given the construction of the variable, they are also more likely to acquire human capital. I find that high PGS respondents were more likely to take additional training while at their first four employers and were more likely to see their primary job duties change while working for an employer. In sum, it seems that high PGSs for educational attainment predict a greater capacity to weather the storms of skill-biased technological change. Some of the heterogeneity in worker responses to retraining programs would therefore seem to be driven by genetics. This has important implications for our understanding of the binding constraints on the success of such programs.

The existing literature raises additional questions of interest to which our results respond.  Barth, Papageorge and Thom (2020) have shown that an educational attainment PGS predicts wealth and retirement, even after controlling for realized education.  Education is inevitably measured with error, not all fields or institutions being equally rigorous.  The educational attainment PGSs' predictive power over and above realized education could be an artefact of differential sorting into majors and universities by PGS.  It could be that high PGS individuals sort into more rigorous educations in young adulthood and reap the benefits of these investments later in their working life.  The significance of a PGS variable even after adjusting for realized education could result from measurement error in the education variable which is correlated with PGSs.  Our finding that an educational attainment PGS predicts economic mobility between the approximate ages of 36 and 53 is evidence against this interpretation.  For early life human capital investments to account for this finding, these investments would have to be very slow to mature.  The results are more consistent with the two stylized models laid out above.   In the first, the genes associated with educational attainment are causally related to economic outcomes throughout the life course and are increasingly predictive of these outcomes with age as the impact of non-shared, non-cumulative environmental shocks is increasingly mean zero with time, a "wash-out" effect.  In the second model, nonshared environmental factors matter, but with age individuals are better positioned to select into environments more suited to their genetically driven endowments and preferences, and the increasing heritability of traits with age reflects an increasing correlation between the direction of genetic and nonshared environmental effects ((Plomin et al., 1977)'s active correlation).  On the basis of the available data, I remain agnostic about the relative importance of these two mechanisms.

Evidence for the genetic basis of important economic outcomes has implications for our understanding of inequality and intergenerational mobility.  Belsky et al. (2018) show that educational attainment PGSs predict educational and wealth mobility relative to parents, and Abdellaoui et al. (2019) show that those with higher educational attainment PGSs have been more likely to move out of Britain's coal mining areas.  Our results are consistent with this literature, and also highlight the importance of measuring economic outcomes at multiple points in time over the life course.  Indeed, genes associated with educational attainment continue to predict economic mobility mid-career.  This suggests that the heritability of economic outcomes is under-estimated if the child's outcomes are measured only in early adulthood, as in some existing studies showing interactions between the PGS and parental socioeconomic status (Houmark, Ronda and Rosholm (2020); Muslimova at al. (2020)).

Outside of the health field, research has paid little attention to possible applications of polygenic scores to economic decision making[7]. An admittedly speculative analogy might be made to "student futures" or income share agreements. If, conditional on typically observable phenotypes, a high educational attainment PGS accurately predicts upward income mobility, it may be in the interest of applicants and lenders of financing for the applicant to disclose genetic data. Indeed, in light of our findings that the educational attainment PGS most strongly predicted upward mobility among those with only a high school education, conceivably a PGS could help to identify "diamonds in the rough" whose phenotypic education reflects a series of negative environmental shocks and are predicted to regress upwards to their long-term genetic mean. It should be noted that larger sample sizes in GWASs used to construct educational attainment PGSs have led to progressively higher shares of phenotypic variance explained, gradually approaching the heritability estimates of twin studies[8]. We are therefore in the early stages of being able to predict complex traits via direct reading of the genome. Polygenic scoring of embryos on the educational attainment variable is already practiced by some firms (Goldberg, 2022), and has generally been viewed as premature (e.g., Lencz et al. 2022), though others have noted that under certain assumptions, even at current accuracy levels this practice may pass a cost benefit test (Shulman and Bostrom, 2014; Branwen, 2016). We can expect potential use of this economically valuable information to be an important social and legal issue in the years to come.

Two additional results of this research suggest that even with current PGSs, genetic disclosure may be welfare enhancing for individuals. Note that the educational attainment PGS predicted mobility most strongly for those graduates with only a high school education, and that the PGS predicted higher aspirations for career mobility only among those with more than a high school education. In tandem, these results suggest that those with a low measured phenotype relative to their genotype early in life may underestimate their potential for economic mobility. Genetic disclosure could be welfare enhancing for these individuals.

The generalizability of these results must be assessed in light of caveats relating to the construction of polygenic scores and the WLS sample. Whether estimated using twin studies or GWASs, heritability estimates are not fixed properties but are estimates drawn from a particular population at a particular point in time. Society-wide environmental changes could change the heritability of traits. For instance, Rimfeld et al. (2018) show that an educational attainment PGS explained twice as much of the variance in educational attainment and economic status in post-Soviet Estonia as in Soviet-era Estonia. In general, an equalization of environments will mean a greater share of phenotypic variation will be accounted for by genetic variation. Similarly, we might be skeptical that heritability estimates derived from contemporary Great Britain or the United States would be replicated in a society rigidly stratified by caste or class, or indeed, across groups within a society characterized by dramatic environmental differences across these groups.

---

[7] Linner and Koellinger (2022) show that PGSs summarizing genetic liability for a series of mortality risks can improve life insurance underwriting, conditional on phenotypes commonly observed during an application for life insurance.

[8] For an introductory discussion of the "missing heritability" problem, see Mills et al. (2020). For recent considerations of the accuracy of polygenic scores from a statistical perspective, see Wang et al. (2022) and Raben et al. (2022).

Moreover, as patterns of linkage disequilibrium differ across ancestral groups, at current GWAS sample sizes studies indicate that PGSs created based on samples from one ancestral group are not as predictive of phenotypes in other ancestral groups (Martin et al., 2017; Duncan et al., 2019). Although there is evidence that the causal alleles explaining variation are held in common across groups (Wang et al., 2022), because of linkage disequilibrium, the SNPs identified by GWASs are not necessarily causal, but may instead be near the causal locus along strands of DNA inherited together. Patterns of linkage disequilibrium differ across ancestral groups. We should therefore be extremely cautious about generalizing these results to populations of non-European descent.

It should also be emphasized that the WLS was designed to be representative of the graduating class of Wisconsin high schools in 1957 and, more broadly, of white American high school seniors of a similar age. Though our sibling-pair estimates convincingly demonstrate the relevance of genetics for economic mobility in this sample, caution should be exercised in generalizing the magnitudes estimated here to the contemporary United States. Indeed, Murnane, Willett and Levy (1995) estimated increasing returns to cognitive abilities across years coincident with our sample, and Papageorge and Thom (2020) find the educational attainment PGS more strongly predicts labour market earnings since 1980. We might therefore expect still larger effect sizes in a regression of economic mobility on polygenic scores for educational attainment in a more contemporary sample.